\documentclass[lettersize,onecolumn]{IEEEtran}

\usepackage{amsmath,amsfonts}
\usepackage{algorithmic}
\usepackage{algorithm}
\usepackage{array}
\usepackage[caption=false,font=normalsize,labelfont=sf,textfont=sf]{subfig}
\usepackage{textcomp}
\usepackage{bigstrut}
\usepackage{multirow}
\usepackage{makecell}
\usepackage{stfloats}
\usepackage{url}
\usepackage{verbatim}
\usepackage{graphicx}
\usepackage[nocompress]{cite}
\usepackage{amsthm}
\usepackage{xcolor}

\newtheorem{theorem}{Theorem}
\newtheorem{lemma}[theorem]{Lemma}
\newtheorem{remark}[theorem]{Remark}

\newtheorem{corollary}[theorem]{Corollary}

\newtheorem{definition}[theorem]{Definition}
\newtheorem{example}[theorem]{Example}

\usepackage[
colorlinks=true,
citecolor=blue,
linkcolor=blue,
urlcolor=blue
]{hyperref}
\begin{document}
	\title{On a class of generalized Roth--Lempel codes
		\thanks{
			This work was supported by the National Natural Science Foundation of China (Grant No. 12471494) and the Natural Science Foundation of Sichuan Province (2024NSFSC2051). The corresponding author is Qunying Liao.
		}
	}
	
	\author{Zhonghao Liang, Yongkang Wan and Qunying Liao* {\thanks{Z. Liang, Y. Wan, and Q. Liao are with the College of Mathematical Science, Sichuan Normal University, Chengdu 610066, China (e-mail: liangzhongh0807@163.com; 2475636261@qq.com; qunyingliao@sicnu.edu.cn)}}
	}
	\maketitle
	
\begin{abstract}
	Maximum distance separable (MDS), almost MDS (AMDS), and
	self-dual codes have applications in communications,
	data storage, and quantum coding.
	The construction and structural study of MDS codes that
	are not equivalent to generalized Reed--Solomon (GRS)
	codes are important problems in coding theory and finite
	geometry.
	In this paper, we investigate the generalized
	Roth--Lempel (GRL) codes $\mathrm{GRL}_{k}(\boldsymbol{\alpha},\boldsymbol{v},\boldsymbol{A}_{3\times 3})$ introduced by Li et al.
	\cite{LFW} in the case of a $3\times3$ appended matrix.
	Using the Cauchy matrix method of Roth and Lempel,
	we derive a sufficient condition under which these codes
	are not equivalent to GRS codes.
	We also establish necessary and sufficient conditions
	for their duals to be AMDS and for the codes to be self-dual.
	The self-duality criterion extends the corresponding
	result in \cite{LFW}, and we provide examples illustrating
	our results.
	Under suitable parameter assumptions and an explicit
	rank condition, we show that the Schur square of an $[n,k]_q$ GRL codes  $\mathrm{GRL}_{k}(\boldsymbol{\alpha},\boldsymbol{v},\boldsymbol{A}_{3\times 3})$  has
	dimension $2k+2$.
	Using this invariant, we show that, under certain
	conditions, these GRL codes are not equivalent
	to codes from several known non-GRS families.
\end{abstract}
	
	
	\section{Introduction}
	\IEEEPARstart{L}{et}  $\mathbb{F}_q$ be the finite field of $q$ elements, where $q$ is a prime power and $\mathbb{F}_{q}^{*}=\mathbb{F}_{q}\backslash\left\{0\right\}$. An $[n, k, d]$ linear code $\mathcal{C}$ over $\mathbb{F}_{q}$ is a $k$-dimensional subspace of $\mathbb{F}_{q}^{n}$ with minimum (Hamming) distance $d$. The dual of an $[n, k]_{q}$ linear code $\mathcal{C}$ is defined as 
	\[
	\mathcal{C}^{\perp}=\left\{\left(x_{1}, \ldots, x_{n}\right)=\boldsymbol{x}\in\mathbb{F}_q^{n} \mid\langle\boldsymbol{x},\boldsymbol{y}\rangle=\sum\limits_{i=1}^{n} x_{i} y_{i}=0,  \forall \boldsymbol{y}=\left(y_{1}, \ldots, y_{n}\right) \in \mathcal{C}\right\}.
	\]
	A linear code $\mathcal{C}$ is self-dual if $\mathcal{C}=\mathcal{C}^{\perp}$. Self-dual linear codes are closely related to combinatorics and lattice theory \cite{A4,A14}. They also have applications in cryptography \cite{A15,A16}. 
	
	The Singleton bound states that $d\leq n-k+1$ for any $[n, k, d]$ linear code $\mathcal{C}$ over $\mathbb{F}_{q}$. If $d=n-k+1$, then $\mathcal{C}$ attains the Singleton bound and is called maximum distance separable (MDS). If $d=n-k$, then $\mathcal{C}$ is called almost MDS (AMDS). If both $\mathcal{C}$ and $\mathcal{C}^{\perp}$ are AMDS, then $\mathcal{C}$ is called near MDS (NMDS). These families have applications in communications, data storage, combinatorics, and secret sharing \cite{A2,A3,A4,A5,A6,A7}. Considerable attention has therefore been devoted to their construction, weight distributions, and equivalence, as well as to self-orthogonality, self-duality, and almost self-duality within these families \cite{A8,A9,A10,A11,A12,A13}. 
	
	Generalized Reed--Solomon (GRS) codes form a fundamental class of MDS linear codes. A GRS code is defined as
	$$
	\mathrm{GRS}_{k}\left(\boldsymbol{\alpha},\boldsymbol{v}\right):=\left\{\left(v_{1}f\left(\alpha_{1}\right), \ldots,v_{n}f\left(\alpha_{n}\right)\right) | f(x) \in \mathbb{F}_{q}^{k}[x]\right\},
	$$
	where $\mathbb{F}_{q}[x]$ is the polynomial ring over $\mathbb{F}_{q}$,
	$$\mathbb{F}_{q}^{k}[x]=\left\{f(x)=\sum_{i=0}^{k-1} f_{i} x^{i}\mid f_{i} \in \mathbb{F}_{q}, 0 \leq i \leq k-1\right\},$$ $\boldsymbol{v}=\left(v_{1}, \ldots, v_{n}\right) \in\left(\mathbb{F}_{q}^{*}\right)^{n}$, and $\boldsymbol{\alpha}=\left(\alpha_{1}, \ldots, \alpha_{n}\right) \in \mathbb{F}_{q}^{n}$ with $\alpha_{i} \neq \alpha_{j}(i \neq j)$. The matrix  
	\begin{equation}\label{RS}
		\boldsymbol{G}_{1}=\left(\begin{matrix}
			v_{1}&		v_{2}&		\cdots&		v_{n-1}&v_{n}\\
			v_{1}\alpha _1&v_{2}\alpha _2&		\cdots&v_{n-1}\alpha _{n-1}&v_{n}\alpha _n\\
			\vdots&		\vdots&		\quad&		\vdots&		\vdots\\
			v_{1}\alpha _{1}^{k-2}&v_{2}\alpha _{2}^{k-2}&		\cdots&v_{n-1}\alpha _{n-1}^{k-2}&v_{n}\alpha _{n}^{k-2}\\
			v_{1}\alpha _{1}^{k-1}&v_{2}\alpha _{2}^{k-1}&		\cdots&v_{n-1}\alpha _{n-1}^{k-1}&v_{n}\alpha _{n}^{k-1}\\ 
		\end{matrix} \right)_{k\times n}
	\end{equation}	
	generates the GRS code $\mathrm{GRS}_{k}(\boldsymbol{\alpha},\boldsymbol{v})$ with parameters $\left[n,k,n-k+1\right]$.
	
A linear code that is equivalent to a GRS code is called a GRS-type code; otherwise, it is called a non-GRS code. The construction and study of non-GRS MDS codes are motivated by the classification of MDS codes and their potential applications in code-based cryptography \cite{A17,A18,A19,A20,A21,A22,LFW,A23,A24,A25,A26}.

In 1989, Roth and Lempel \cite{A21} constructed a class of non-GRS MDS codes by appending two columns to the matrix $\boldsymbol{G}_{1}$ in $(\ref{RS})$. The resulting linear codes over $\mathbb{F}_{q}$ have a generator matrix
\begin{equation}\label{G_2}
	\boldsymbol{G}_{2}
	=\left(
	\begin{matrix}
		\boldsymbol{G}_{1}&
		\begin{matrix}
			\boldsymbol{0}\\
			\boldsymbol{A}_1
		\end{matrix}
	\end{matrix}
	\right)_{k\times(n+2)},
\end{equation}
where $4\leq k+1\leq n\leq q$ and
$\boldsymbol{A}_1=\left(\begin{matrix}
	0&1\\
	1&\delta
\end{matrix}\right)$ with $\delta\in\mathbb{F}_{q}$.

More recently, Wu et al. \cite{A22} studied a construction obtained by appending three columns to $\boldsymbol{G}_{1}$. The resulting linear codes over $\mathbb{F}_{q}$ have a generator matrix
\begin{equation}\label{G_3}
	\boldsymbol{G}_{3}
	=\left(
	\begin{matrix}
		\boldsymbol{G}_{1}&
		\begin{matrix}
			\boldsymbol{0}\\
			\boldsymbol{A}_2
		\end{matrix}
	\end{matrix}
	\right)_{k\times(n+3)},
\end{equation}
where $4\leq k+1\leq n\leq q$ and
$\boldsymbol{A}_2=\left(\begin{matrix}
	0&0&1\\
	0&1&\tau\\
	1&\delta&\mu
\end{matrix}\right)$ with $\delta,\tau,\mu\in\mathbb{F}_{q}$.
They used code extensions to prove that these codes are not equivalent to GRS codes. They also established necessary and sufficient conditions for the codes or their duals to be MDS, AMDS, or NMDS, determined the covering radii and deep holes of the duals for a particular subfamily, and obtained an infinite family of three-dimensional codes that are optimally or almost optimally extendable.

Li et al. \cite{LFW} subsequently introduced generalized Roth--Lempel (GRL) codes, providing a common framework for these constructions. They determined a parity-check matrix and established necessary and sufficient conditions for two classes of GRL codes to be MDS or self-dual, respectively. Their self-duality result assumes $q\equiv1\pmod4$.

The GRL construction allows the appended matrix to vary. It is therefore natural to investigate how the choice of this matrix affects the properties of the resulting codes. In particular, conditions for the codes to be non-GRS, for their duals to be AMDS, and for the codes themselves to be self-dual help identify matrices that yield the desired properties. For self-duality, the assumption $q\equiv1\pmod4$ in \cite{LFW} also leads to the question of whether self-dual GRL codes can be obtained without this restriction.

Another question concerns the relation between GRL codes and other known non-GRS codes. A code that is not equivalent to a GRS code may still be equivalent to a code from another known family. Beelen et al. \cite{2018TGRS} introduced twisted Reed--Solomon codes,  studied their structural properties and their resistance to certain known attacks on Reed--Solomon-based cryptosystems, including the Sidelnikov--Shestakov and Wieschebrink attacks. They also compared twisted Reed--Solomon codes with classical Roth--Lempel codes computationally and found that the two families are largely independent for the parameters considered.

More recently, Hu et al. \cite{HuTGRSTIT2025} introduced $(\mathcal L,\mathcal P)$-twisted generalized Reed--Solomon codes, providing a general framework for twisted Reed--Solomon constructions . Thus, both the classical Roth--Lempel and twisted Reed--Solomon constructions have been generalized. It is natural to ask whether similar distinctions can be established between their generalized forms. This motivates us to investigate conditions under which GRL codes are not equivalent to specific classes of $(\mathcal L,\mathcal P)$-twisted generalized Reed--Solomon codes, as well as other known non-GRS codes.

	Motivated by these questions, we consider GRL codes obtained by replacing $\boldsymbol{A}_{i}(i=1,2)$ with $\boldsymbol{A}_{3\times 3}=(a_{ij})\in\mathrm{GL}_{3}\left(\mathbb{F}_{q}\right)$. First, we derive a sufficient condition for these GRL codes to be non-GRS when $k>4$ using the Cauchy matrix method of Roth and Lempel \cite{A21,RLCauchy}. Second, we establish a necessary and sufficient condition for their duals to be AMDS. Third, we establish a necessary and sufficient condition for self-duality. Finally, by computing Schur square dimensions, we show that, under certain conditions, these GRL codes are not monomially equivalent to several known classes of non-GRS linear codes.
	
	The remainder of this paper is organized as follows. In Section \ref{sec 2}, we recall the definition of GRL codes and present the preliminary lemmas. In Section \ref{sec 3}, we establish a sufficient condition for $\mathrm{GRL}_{k}(\boldsymbol{\alpha},\boldsymbol{v},\boldsymbol{A}_{3\times 3})(k>4)$ to be non-GRS. In Section \ref{sec 4}, we establish a necessary and sufficient condition for the dual of $\mathrm{GRL}_{k}(\boldsymbol{\alpha},\boldsymbol{v},\boldsymbol{A}_{3\times 3})$ to be AMDS. In Section \ref{sec 5}, using a parity-check matrix of $\mathrm{GRL}_{k}(\boldsymbol{\alpha},\boldsymbol{v},\boldsymbol{A}_{3\times 3})$, we derive a necessary and sufficient condition for self-duality. In Section~\ref{sec 6}, we compare these GRL codes with several known non-GRS code families using Schur square dimensions. In Section \ref{sec 7}, we conclude the paper. 
	
	\section{Preliminaries}\label{sec 2}
	In this section, we recall the definition of GRL codes and present the preliminary lemmas.
	
	In 2025, Li et al. \cite{LFW} introduced GRL codes as follows.
	\begin{definition}\label{definition1}{\rm(\cite{LFW}, Definition 2.3)}
		Let $\mathbb{F}_q$ be the finite field with $q$ elements, where $q$ is a prime power. Let $l+1\leq k+1\leq n\leq q$,  $\boldsymbol{\alpha}=\left(\alpha_{1}, \ldots, \alpha_{n}\right) \in \mathbb{F}_{q}^{n}$ with $\alpha_{i} \neq \alpha_{j}(i \neq j)$ and $\boldsymbol{v}=$ $\left(v_{1}, \ldots, v_{n}\right) \in\left(\mathbb{F}_{q}^{*}\right)^{n}$. The generalized Roth--Lempel (GRL) code $\mathrm{GRL}_{k}(\boldsymbol{\alpha}, \boldsymbol{v},\boldsymbol{A}_{l\times l})$ is defined by
		$$
		\mathrm{GRL}_{k}(\boldsymbol{\alpha}, \boldsymbol{v},\boldsymbol{A}_{l\times l}):=\left\{\left(v_{1} f\left(\alpha_{1}\right), \ldots, v_{n} f\left(\alpha_{n}\right),\boldsymbol{\beta}\right) | f(x) \in \mathbb{F}_{q}^{k}[x]\right\},
		$$
		where $\boldsymbol{A}_{l\times l}=(a_{ij})_{l\times l}\in\mathrm{GL}_{l}\left(\mathbb{F}_{q}\right)$,
		$$\boldsymbol{\beta}=\left(f_{k-l},\ldots,f_{k-1}\right)\boldsymbol{A}_{l\times l}=\left(a_{11}f_{k-l}+a_{21}f_{k-(l-1)}+\cdots+a_{l1}f_{k-1},\ldots,a_{1l}f_{k-l}+a_{2l}f_{k-(l-1)}+\cdots+a_{ll}f_{k-1}\right),$$
		and $f_t$ denotes the coefficient of $x^{t}$ in $f(x)$.
	\end{definition} 
	\begin{remark}\label{remark1}
		(1) Taking $\boldsymbol{v}=\left(1, \ldots, 1\right)$ and $\boldsymbol{A}_{l\times l}=\boldsymbol{A}_1$ in Definition $\ref*{definition1}$, we obtain $\mathrm{GRL}_{k}(\boldsymbol{\alpha}, \boldsymbol{v},\boldsymbol{A}_{l\times l})$, the classical Roth--Lempel (RL) code denoted by $\mathrm{RL}_{k}(\boldsymbol{\alpha},\delta)$.
		
		(2) Taking $\boldsymbol{v}=\left(1, \ldots, 1\right)$ and $\boldsymbol{A}_{l\times l}=\boldsymbol{A}_{3\times 3}$ in Definition $\ref*{definition1}$, we obtain a code with generator matrix \begin{equation}\label{RL}
			\boldsymbol{G}_{4}= \begin{pmatrix}
				1&		1&		\cdots&		1&1&0&0&0\\
				\alpha _1&\alpha _2&		\cdots&\alpha _{n-1}&\alpha _n&0&0&0\\
				\vdots&		\vdots& &		\vdots&		\vdots&\vdots&\vdots&\vdots\\
				\alpha _{1}^{k-4}&\alpha _{2}^{k-4}&		\cdots&\alpha _{n-1}^{k-4}&\alpha _{n}^{k-4}&0&0&0\\
				\alpha _{1}^{k-3}&\alpha _{2}^{k-3}&		\cdots&\alpha _{n-1}^{k-3}&\alpha _{n}^{k-3}&a_{11}&a_{12}&a_{13}\\
				\alpha _{1}^{k-2}&\alpha _{2}^{k-2}&		\cdots&\alpha _{n-1}^{k-2}&\alpha _{n}^{k-2}&a_{21}&a_{22}&a_{23}\\
				\alpha _{1}^{k-1}&\alpha _{2}^{k-1}&		\cdots&\alpha _{n-1}^{k-1}&\alpha _{n}^{k-1}&a_{31}&a_{32}&a_{33}\\ 
			\end{pmatrix}_{k\times (n+3)}.
		\end{equation} 
	\end{remark}  
	Taking $m=n-3$,  $(k_1,\ldots,k_{m-1},k_m)=(1,\ldots,n-4,n-1)$ and $(x_1,\ldots,x_{m+1})=(\alpha_1,\ldots,\alpha_{n-2})$ in \cite[Lemma 2]{A22} gives Lemma~\ref{lemma1} below.
	\begin{lemma}\label{lemma1}{\rm(\cite{A22}, Lemma 2)} For every integer $n>3$, we have 
		$$\det\begin{pmatrix}
			1 & 1 & \ldots & 1 \\
			\alpha_{1} & \alpha_{2} & \ldots & \alpha_{n-2} \\
			\vdots & \vdots & \ddots & \vdots \\
			\alpha_{1}^{n-4} & \alpha_{2}^{n-4} & \ldots & \alpha_{n-2}^{n-4} \\
			\alpha_{1}^{n-1} & \alpha_{2}^{n-1} & \ldots & \alpha_{n-2}^{n-1}
		\end{pmatrix}=\left(\sum_{i=1}^{n-2} \alpha_{i}^{2}+\sum_{1\leq i<j \leq n-2} \alpha_{i} \alpha_{j}\right)\prod_{1 \leq i<j \leq n-2}\left(\alpha_{j}-\alpha_{i}\right) .$$
	\end{lemma} 
	
	\begin{lemma}\label{lemma4}
		{\rm(\cite{A29}, Proposition 2.1)} An $[n, k]$ linear code over $\mathbb{F}_{q}$ is $\mathrm{MDS}$ if and only if any $k$ columns of its generator matrix are $\mathbb{F}_{q}$-linearly independent.	
	\end{lemma}
	
	\begin{lemma}\label{monomiallyequivalent}{\rm(\cite{A3}, Page 24, Section 1.7)}
		Let $\mathcal{C}_{1}$ and $\mathcal{C}_{2}$ be two linear codes  of the same length over $\mathbb{F}_{q}$, and let ${\color{red} \boldsymbol{G}}$ be a generator matrix of $\mathcal{C}_{1}$. Then $\mathcal{C}_{1}$ and $\mathcal{C}_{2}$ are monomially equivalent if and only if there exists a monomial matrix $\boldsymbol{D}$ such that $\boldsymbol{{\color{red}G}D}$ is a generator matrix of $\mathcal{C}_{2}$.
	\end{lemma}
	
	The following two lemmas relate GRS codes to generalized Cauchy matrices and characterize the latter through their entrywise reciprocals. These results will be used to establish the non-GRS property.
	\begin{lemma}\label{lem:wang-cauchy}{\rm(\cite{A21}; \cite{WangMGRS}, Lemma 1)}
		A matrix of the form $\begin{pmatrix}
			\boldsymbol I_k&\boldsymbol F
		\end{pmatrix}$ with $\boldsymbol F=(\boldsymbol F_{ij})$
		generates a $[n,k]_q$ GRS code if and only if $\boldsymbol F$
		is a generalized Cauchy matrix, i.e., 
		$$\boldsymbol F_{ij}
		=
		\frac{c_i d_j}{x_i+y_j}, 0\leq i\leq k-1,0\leq j\leq n-k-1,$$
		where the $x_i$ are distinct elements of $\mathbb{F}_q$, the $y_i$ are distinct
		elements of $\mathbb{F}_q$, $x_i+y_j\neq 0$ for all $i$ and $j$, and $c_i,d_j\neq 0$.
	\end{lemma} 
	\begin{lemma}\label{lem:wang-reciprocal}{\rm(\cite{A21,A22}; \cite{WangMGRS}, Lemma 2)}
		Let $\mathcal{C}$ be a $[n,k]_q$ linear code with generator matrix
		of the form $\begin{pmatrix}
			\boldsymbol I_k&\boldsymbol F
		\end{pmatrix}$ with $\boldsymbol F=(\boldsymbol F_{ij}),0\leq i\leq k-1,0\leq j\leq n-k-1$.
		Then \(\boldsymbol F\) is a generalized Cauchy matrix if and only if the following three conditions hold simultaneously.
		\begin{enumerate}
			\item All entries of \(\boldsymbol F\) are nonzero;
			\item All \(2\times2\) minors of \(\boldsymbol F^\sharp
			=
			\left(\frac{1}{\boldsymbol F_{ij}}\right)\) are nonzero;
			\item All \(3\times3\) minors of \(\boldsymbol F^\sharp
			=
			\left(\frac{1}{\boldsymbol F_{ij}}\right)\) are zero.
		\end{enumerate}
	\end{lemma}
	\section{The non-GRS property of $\mathrm{GRL}_{k}(\boldsymbol{\alpha},\boldsymbol{v},\boldsymbol{A}_{3\times 3})$}\label{sec 3}
	In this section, we derive a sufficient condition for $\mathrm{GRL}_{k}(\boldsymbol{\alpha},\boldsymbol{v},\boldsymbol{A}_{3\times 3})(k>4)$ to be non-GRS.
	
	We begin with a lemma on the inverse of a Vandermonde matrix.
	\begin{lemma}
		\label{lem:vandermonde-inverse}
		Let \(\alpha_1,\ldots,\alpha_k\) be pairwise distinct elements of
		\(\mathbb F_q\), and define
		\[
		\boldsymbol V_k=
		\begin{pmatrix}
			1&1&\cdots&1\\
			\alpha_1&\alpha_2&\cdots&\alpha_k\\
			\vdots&\vdots&&\vdots\\
			\alpha_1^{k-1}&\alpha_2^{k-1}&\cdots&\alpha_k^{k-1}
		\end{pmatrix}.
		\]
		For each \(1\le i\le k\), define
		\[
		f_i(x)=\prod_{\substack{1\le j\le k\\j\ne i}}(x-\alpha_j),
		\qquad
		\eta_i=f_i(\alpha_i),
		\]
		and
		\[
		L_i(x)=\frac{f_i(x)}{\eta_i}
		=\sum_{r=0}^{k-1}\ell_{i,r}x^r.
		\]
		Then
		\[
		(\boldsymbol V_k^{-1})_{i,r+1}=\ell_{i,r},
		\qquad 0\le r\le k-1.
		\]
	\end{lemma}
	
	\begin{proof}
		For each \(1\le i\le k\), let $\boldsymbol\ell_i^T
		=
		\begin{pmatrix}
			\ell_{i,0}&\ell_{i,1}&\cdots&\ell_{i,k-1}
		\end{pmatrix}.$ By the definition of the Lagrange polynomials, $$L_i(\alpha_j)=\begin{cases}
			1,& i=j;\\
			0, & i\neq j,
		\end{cases}$$ and hence 
		\[
		\boldsymbol\ell_i^T\boldsymbol V_k
		=
		\begin{pmatrix}
			L_i(\alpha_1)&L_i(\alpha_2)&\cdots&L_i(\alpha_k)
		\end{pmatrix}
		=
		\begin{pmatrix}
			0&\cdots&0&1&0&\cdots&0
		\end{pmatrix},
		\]
		where the entry \(1\) is in the \(i\)-th position. Multiplying both sides on the right by \(\boldsymbol V_k^{-1}\) gives
		\[
		\boldsymbol\ell_i^T
		=
		\begin{pmatrix}
			0&\cdots&0&1&0&\cdots&0
		\end{pmatrix}
		\boldsymbol V_k^{-1}.
		\]
		Hence \(\boldsymbol\ell_i^T\) is the \(i\)-th row of
		\(\boldsymbol V_k^{-1}\). Therefore, the $(i,r+1)$-th entry $(\boldsymbol V_k^{-1})_{i,r+1}$ of $\boldsymbol V_k^{-1}$ is $
		\ell_{i,r}$ for $0\le r\le k-1.$
	\end{proof} 
	\begin{lemma}\label{lem:systematic-form}
		Retain the notation of Lemma~\ref{lem:vandermonde-inverse}. Define
		\[
		\sigma_1=\sum_{j=1}^k\alpha_j,
		\qquad
		\sigma_2=\sum_{1\le r<t\le k}\alpha_r\alpha_t,
		\]
		and, for \(s=1,2,3\),
		\[
		P_s(x)
		=
		a_{3s}-a_{2s}\sigma_1+a_{1s}\sigma_2
		+(a_{2s}-a_{1s}\sigma_1)x
		+a_{1s}x^2.
		\]
		Then
		\[
		\boldsymbol V_k^{-1}\boldsymbol G_4
		=
		\begin{pmatrix}
			\boldsymbol I_k&\boldsymbol F
		\end{pmatrix},
		\]
		where $\boldsymbol F=\left(\boldsymbol F_{ij}\right)_{1\leq i\leq k,k+1\leq j\leq n+3}$ with $\boldsymbol F_{ij}=\begin{cases}
			\frac{f_i(\alpha_j)}{\eta_i},& 1\leq i\leq k,k+1\le j\le n;\\
			\frac{P_s(\alpha_i)}{\eta_i}&1\leq i\leq k,j=n+s,s=1,2,3.
		\end{cases}$.
	\end{lemma}
	
	\begin{proof}
		For \(k+1\le j\le n\), the \(j\)-th column of \(\boldsymbol G_4\) is
		\[
		(1,\alpha_j,\alpha_j^2,\ldots,\alpha_j^{k-1})^T.
		\]
		Thus, by Lemma~\ref{lem:vandermonde-inverse}, we have
		$$\boldsymbol F_{ij}=\sum_{r=0}^{k-1}
		(\boldsymbol V_k^{-1})_{i,r+1}\alpha_j^r=\sum_{r=0}^{k-1}\ell_{i,r}\alpha_j^r=L_i(\alpha_j)=\frac{f_i(\alpha_j)}{\eta_i}(1\leq i\leq k, k+1\le j\le n),$$ 
		and 
		$$
		\boldsymbol F_{i(n+s)}
		=
		a_{1s}\ell_{i,k-3}
		+a_{2s}\ell_{i,k-2}
		+a_{3s}\ell_{i,k-1}( 1\leq i\leq k,s=1,2,3).
		$$
		Note that  
		\[
		f_i(x)
		=
		x^{k-1}
		-\left(\sum_{j\ne i}\alpha_j\right)x^{k-2}
		+\left(\sum_{\substack{r<t\\r,t\ne i}}
		\alpha_r\alpha_t\right)x^{k-3}
		+\cdots=x^{k-1}
		-\left(\sigma_1-\alpha_i\right)x^{k-2}
		+\left(
		\sigma_2-\sigma_1\alpha_i+\alpha_i^2\right)x^{k-3}
		+\cdots,
		\]  
		Comparing coefficients gives
		$$
		\ell_{i,t}
		=\begin{cases} 
			\frac{\sigma_2-\sigma_1\alpha_i+\alpha_i^2}{\eta_i},& t=k-3;\\
			\frac{\alpha_i-\sigma_1}{\eta_i},& t=k-2;\\
			\frac{1}{\eta_i},& t=k-1.
		\end{cases}$$
		Hence,
		$$
		\boldsymbol F_{i(n+s)}=
		\frac{
			a_{3s}-a_{2s}\sigma_1+a_{1s}\sigma_2
		}{\eta_i}
		+
		\frac{
			(a_{2s}-a_{1s}\sigma_1)\alpha_i
			+a_{1s}\alpha_i^2
		}{\eta_i}=
		\frac{P_s(\alpha_i)}{\eta_i}.$$
		This proves the lemma.
	\end{proof}   
	We now use Lemmas~\ref{lem:vandermonde-inverse} and \ref{lem:systematic-form} to prove the following theorem.
	\begin{theorem}
		\label{thm:non-GRS}
		Let  
		\begin{equation}\label{nonGRSA}
			\boldsymbol{W}=
			\begin{pmatrix}
				p_{20}p_{30}&
				p_{20}p_{31}+p_{21}p_{30}&
				p_{20}p_{32}+p_{21}p_{31}+p_{22}p_{30}&
				p_{21}p_{32}+p_{22}p_{31}&
				p_{22}p_{32}\\[3pt]
				p_{10}p_{30}&
				p_{10}p_{31}+p_{11}p_{30}&
				p_{10}p_{32}+p_{11}p_{31}+p_{12}p_{30}&
				p_{11}p_{32}+p_{12}p_{31}&
				p_{12}p_{32}\\[3pt]
				p_{10}p_{20}&
				p_{10}p_{21}+p_{11}p_{20}&
				p_{10}p_{22}+p_{11}p_{21}+p_{12}p_{20}&
				p_{11}p_{22}+p_{12}p_{21}&
				p_{12}p_{22}
			\end{pmatrix},
		\end{equation}
		where
		\[
		\begin{aligned}
			p_{s0}&=a_{3s}-a_{2s}\sigma_1+a_{1s}\sigma_2,\\
			p_{s1}&=a_{2s}-a_{1s}\sigma_1,\\
			p_{s2}&=a_{1s}.
		\end{aligned}
		\]
		If $\mathrm{rank}\left(\boldsymbol{W}\right)=3$, then $\mathrm{GRL}_k(\boldsymbol\alpha,\boldsymbol v,
		\boldsymbol A_{3\times3})$
		is non-GRS for $k\geq 5$.
	\end{theorem}
	
	\begin{proof}
		By Lemma~\ref{monomiallyequivalent}, it suffices to consider $\mathrm{GRL}_k(\boldsymbol\alpha,\boldsymbol 1,
		\boldsymbol A_{3\times3})$. By Lemma~\ref{lem:systematic-form}, $\mathrm{GRL}_k(\boldsymbol\alpha,\boldsymbol 1,
		\boldsymbol A_{3\times3})$  has a generator matrix $$
		\boldsymbol V_k^{-1}\boldsymbol G_4
		=
		\begin{pmatrix}
		\boldsymbol I_k&\boldsymbol F
		\end{pmatrix},$$
		where
		\[
		\boldsymbol F_{ij}=\frac{f_i(\alpha_j)}{\eta_i}( 1\leq i\leq k,k+1\le j\le n),
		\]
		and
		\[
		\boldsymbol F_{i(n+s)}=\frac{P_s(\alpha_i)}{\eta_i}( 1\leq i\leq k,s=1,2,3).
		\]
		
		Suppose that $\mathrm{GRL}_k(\boldsymbol\alpha,\boldsymbol 1,
		\boldsymbol A_{3\times3})$ is a $[n+3,k]_q$ GRS code. By Lemma~\ref{lem:wang-cauchy}, \(\boldsymbol F\) is a generalized Cauchy matrix with $F_{i(n+s)}\neq 0$. Furthermore, by Lemma~\ref{lem:wang-reciprocal}, all \(3\times3\) minors of \(\boldsymbol F^\sharp
		=
		\left(\frac{1}{\boldsymbol F_{ij}}\right)\) are zero. Moreover, $F_{ij}\neq 0$ implies that $P_{s}\left(\alpha_{i}\right)\neq 0$.
		
		We obtain a contradiction from the following claim.
		
		{\bf Claim A} If $\mathrm{rank}\left(\boldsymbol{W}\right)=3$, then there exists a \(3\times3\) nonzero minor of \(\boldsymbol F^\sharp
		=
		\left(\frac{1}{\boldsymbol F_{ij}}\right)\).
		
		To prove {\bf Claim A}, it suffices to show that the last three columns of \(\boldsymbol F^\sharp
		=
		\left(\frac{1}{\boldsymbol F_{ij}}\right)\) are $\mathbb{F}_q$-linearly independent when $\mathrm{rank}\left(\boldsymbol{W}\right)=3$.
		
		Otherwise, if there exists a tuple $\left(\lambda_1,\lambda_2,\lambda_3\right)\in\mathbb{F}_{q}^3\backslash\left\{\boldsymbol{0}\right\}$ such that, for every $i$ with \(1\le i\le k\),
		\[
		\lambda_1\frac{1}{\boldsymbol F_{i(n+1)}}
		+
		\lambda_2\frac{1}{\boldsymbol F_{i(n+2)}}
		+
		\lambda_3\frac{1}{\boldsymbol F_{i(n+3)}}
		=0.
		\]
		Since $\frac{1}{\boldsymbol F_{i(n+s)}}
		=
		\frac{\eta_i}{P_s(\alpha_i)},$ $\eta_i\ne0$ and $P_{s}\left(\alpha_{i}\right)\neq 0$,
		the polynomial  
		\[
		R(x)
		=
		\lambda_1P_2(x)P_3(x)
		+
		\lambda_2P_1(x)P_3(x)
		+
		\lambda_3P_1(x)P_2(x)
		\] 
		has the $k\geq 5$ distinct roots $\alpha_{1},\ldots,\alpha_{k}$ in $\mathbb{F}_{q}$. Since $$
		P_s(x)
		=
		a_{3s}-a_{2s}\sigma_1+a_{1s}\sigma_2
		+(a_{2s}-a_{1s}\sigma_1)x
		+a_{1s}x^2
		,$$ it follows that $\deg \left(P_s(x)\right)\le2,$
		so
		\[
		\deg R(x)\le4,
		\]
		so the tuple $\left(\lambda_1,\lambda_2,\lambda_3\right)\in\mathbb{F}_{q}^3\backslash\left\{\boldsymbol{0}\right\}$ satisfies $R(x)=0$. On the other hand, comparing coefficients gives
		$$R(x)=0\Longleftrightarrow \begin{pmatrix}
			p_{20}p_{30}&
			p_{20}p_{31}+p_{21}p_{30}&
			p_{20}p_{32}+p_{21}p_{31}+p_{22}p_{30}&
			p_{21}p_{32}+p_{22}p_{31}&
			p_{22}p_{32}\\[3pt]
			p_{10}p_{30}&
			p_{10}p_{31}+p_{11}p_{30}&
			p_{10}p_{32}+p_{11}p_{31}+p_{12}p_{30}&
			p_{11}p_{32}+p_{12}p_{31}&
			p_{12}p_{32}\\[3pt]
			p_{10}p_{20}&
			p_{10}p_{21}+p_{11}p_{20}&
			p_{10}p_{22}+p_{11}p_{21}+p_{12}p_{20}&
			p_{11}p_{22}+p_{12}p_{21}&
			p_{12}p_{22}
		\end{pmatrix}^T\begin{pmatrix}
			\lambda_1\\
			\lambda_2\\
			\lambda_3
		\end{pmatrix}=0,$$
		i.e., $\boldsymbol{W}^T\begin{pmatrix}
			\lambda_1\\
			\lambda_2\\
			\lambda_3
		\end{pmatrix}=0$. Since $\mathrm{rank}\left(\boldsymbol{W}^T\right)=3$, the system $\boldsymbol{W}^T\begin{pmatrix}
			\lambda_1\\
			\lambda_2\\
			\lambda_3
		\end{pmatrix}=0$ has only the zero solution. Thus, when $\mathrm{rank}\left(\boldsymbol{W}^T\right)=3$, for every $\left(\lambda_1,\lambda_2,\lambda_3\right)\in\mathbb{F}_{q}^3\backslash\left\{\boldsymbol{0}\right\}$, we have $R(x)\neq 0$, a contradiction. Consequently, when $\mathrm{rank}\left(\boldsymbol{W}\right)=3$, the last three columns of \(\boldsymbol F^\sharp
		=
		\left(\frac{1}{\boldsymbol F_{ij}}\right)\) are $\mathbb{F}_q$-linearly independent, and hence there exists a \(3\times3\) nonzero minor of \(\boldsymbol F^\sharp
		=
		\left(\frac{1}{\boldsymbol F_{ij}}\right)\).
		Hence, when $\mathrm{rank}\left(\boldsymbol{W}\right)=3$, $\mathrm{GRL}_k(\boldsymbol\alpha,\boldsymbol v,
		\boldsymbol A_{3\times3})
		$
		is non-GRS for \(k\ge5\).
	\end{proof}
	\begin{theorem}
		\label{tm:R-nonzero-triangular}
		If $\boldsymbol A_{3\times3}
		=
		\begin{pmatrix}
			0&0&a_{13}\\
			0&a_{22}&a_{23}\\
			a_{31}&a_{32}&a_{33}
		\end{pmatrix}\in\mathrm{GL}_{3}\left(\mathbb{F}_{q}\right),$ then, for every  $\left(\lambda_1,\lambda_2,\lambda_3\right)\in\mathbb{F}_{q}^3\backslash\left\{\boldsymbol{0}\right\}$, we have $$R(x)
		=
		\lambda_1P_2(x)P_3(x)
		+
		\lambda_2P_1(x)P_3(x)
		+
		\lambda_3P_1(x)P_2(x)\neq 0.$$ Furthermore, the corresponding GRL code $\mathrm{GRL}_k(\boldsymbol\alpha,\boldsymbol v,
		\boldsymbol A_{3\times3})
		$
		is non-GRS for \(k\ge5\).
	\end{theorem}
	
	\begin{proof}
		Since $\boldsymbol A_{3\times3}
		=
		\begin{pmatrix}
			0&0&a_{13}\\
			0&a_{22}&a_{23}\\
			a_{31}&a_{32}&a_{33}
		\end{pmatrix}\in\mathrm{GL}_{3}\left(\mathbb{F}_{q}\right),$ we have $a_{11}=a_{12}=a_{21}=0$ and $a_{13},a_{22},a_{31}\in\mathbb{F}_{q}^{*}$. Consequently,
		$$P_1(x)=a_{31},$$ 
		$$P_2(x)
		=
		a_{32}-a_{22}\sigma_1+a_{22}x,
		$$
		and
		$$
		P_3(x)
		=
		a_{33}-a_{23}\sigma_1+a_{13}\sigma_2
		+(a_{23}-a_{13}\sigma_1)x
		+a_{13}x^2,
		$$  
		Therefore, 
		$$\deg\left(P_{2}(x)P_{3}(x)\right)=3,\deg\left(P_{1}(x)P_{3}(x)\right)=2,\deg\left(P_{1}(x)P_{2}(x)\right)=1.$$	
		Moreover, their leading coefficients $a_{22}a_{13}, a_{13}a_{31}, a_{31}a_{22}$	
		are all nonzero. Thus $P_2(x)P_3(x),P_1(x)P_3(x),P_1(x)P_2(x)$ are $\mathbb{F}_{q}$-linearly independent, so \(R(x)\neq0\) for every  $\left(\lambda_1,\lambda_2,\lambda_3\right)\in\mathbb{F}_{q}^3\backslash\left\{\boldsymbol{0}\right\}$. The non-GRS conclusion follows by the same argument as in the proof of Theorem \ref{thm:non-GRS}.
	\end{proof}
	\begin{remark}
		Wu et al. \cite{A22} proved that the corresponding code $\mathrm{GRL}_k(\boldsymbol\alpha,\boldsymbol v,
		\boldsymbol A_{3\times3})
		$
		is non-GRS for $\boldsymbol A_{3\times3}
		=
		\begin{pmatrix}
			0&0&1\\
			0&1&a_{23}\\
			1&a_{32}&a_{33}
		\end{pmatrix}\in\mathrm{GL}_{3}\left(\mathbb{F}_{q}\right)$ using code extensions. Theorem \ref{tm:R-nonzero-triangular} provides an alternative proof.
	\end{remark}
	\section{A necessary and sufficient condition for $\mathrm{GRL}_{k}^{\perp}(\boldsymbol{\alpha},\boldsymbol{v},\boldsymbol{A}_{3\times 3})$ to be AMDS }\label{sec 4}
	In this section, we establish a necessary and sufficient condition for $\mathrm{GRL}_{k}^{\perp}(\boldsymbol{\alpha},\boldsymbol{v},\boldsymbol{A}_{3\times 3})$ to be AMDS.
	
	We first recall a necessary and sufficient condition for $\mathrm{GRL}_{k}(\boldsymbol{\alpha},\boldsymbol{v},\boldsymbol{A}_{3\times 3})$ to be MDS.  
	\begin{lemma}\label{MDSRLCODES}{\rm(\cite{LFW}, Definition 2.3)}
		If $k\geq 4$, then $\mathrm{GRL}_{k}(\boldsymbol{\alpha},\boldsymbol{v},\boldsymbol{A}_{3\times 3})$ is $\mathrm{MDS}$ if and only if the following two conditions hold simultaneously:
		
		$(1)$ for every subset $J\subseteq\left\{\alpha _{1},\ldots,\alpha _{n}\right\}$ of size $k-1$,
		$$a_{1s}\sum\limits_{\alpha_{i_{l}},\alpha_{i_{j}}\in J,i_{l}<i_{j}}\alpha_{i_{l}}\alpha_{i_{j}}+a_{3s}\neq a_{2s}\sum\limits_{\alpha_{i_{l}}\in J}\alpha_{i_{l}}(1\leq s\leq 3);$$
		
		$(2)$ for every subset $I\subseteq\left\{\alpha _{1},\ldots,\alpha _{n}\right\}$ of size $k-2$,
		$$\left(-a_{1s}a_{2t}+a_{2s}a_{1t}\right)\left(\sum\limits_{\alpha_{i_{l}}\in I}\alpha_{i_{l}}^{2}+\sum\limits_{\alpha_{i_{l}},\alpha_{i_{j}}\in I,i_{l}<i_{j}}\alpha_{i_{l}}\alpha_{i_{j}}\right)\neq\left(a_{3s}a_{1t} -a_{1s}a_{3t}\right)\sum\limits_{\alpha_{i_{l}}\in I}\alpha_{i_{l}}+a_{2s}a_{3t}-a_{3s}a_{2t}(1\leq t<s\leq 3).$$
	\end{lemma} 
	
	Using Lemma~\ref{MDSRLCODES}, we obtain the following criterion.
	\begin{theorem}\label{AMDSDUALRLCODES}
		If $k\geq 4$, then $\mathrm{GRL}_{k}^{\perp}(\boldsymbol{\alpha},\boldsymbol{v},\boldsymbol{A}_{3\times 3})$ is AMDS if and only if the following conditions hold simultaneously:
		
		$(1)$ for every subset $I\subseteq\left\{\alpha _{1},\ldots,\alpha _{n}\right\}$ of size $k-2$ and every $1\leq r\leq 3$, at least one of the following conditions holds:
		$$a_{2r}\neq a_{1r}\sum\limits_{\alpha _{i_l}\in I}\alpha _{i_l},\  a_{3r}\neq a_{1r}\left(\sum\limits_{\alpha_{i_{l}}\in I}\alpha_{i_{l}}^{2}+\sum\limits_{\alpha_{i_{l}},\alpha_{i_{j}}\in I,i_{l}<i_{j}}\alpha_{i_{l}}\alpha_{i_{j}}\right),\ a_{3r}\sum\limits_{\alpha _{i_l}\in I}\alpha _{i_l}\neq a_{2r}\left(\sum\limits_{\alpha_{i_{l}}\in I}\alpha_{i_{l}}^{2}+\sum\limits_{\alpha_{i_{l}},\alpha_{i_{j}}\in I,i_{l}<i_{j}}\alpha_{i_{l}}\alpha_{i_{j}}\right);$$ 
		
		$(2)$ at least one of the following conditions holds:
		
		$(2.1)$ there exists a subset $J\subseteq\left\{\alpha _{1},\ldots,\alpha _{n}\right\}$ of size $k-1$ such that at least one of the following conditions holds:
		$$a_{1s}\sum\limits_{\alpha_{i_{l}},\alpha_{i_{j}}\in J,i_{l}<i_{j}}\alpha_{i_{l}}\alpha_{i_{j}}+a_{3s}= a_{2s}\sum\limits_{\alpha_{i_{l}}\in J}\alpha_{i_{l}}(1\leq s\leq 3);$$
		
		$(2.2)$ there exists a subset $I\subseteq\left\{\alpha _{1},\ldots,\alpha _{n}\right\}$ of size $k-2$ such that at least one of the following conditions holds:
		$$\left(-a_{1s}a_{2t}+a_{2s}a_{1t}\right)\left(\sum\limits_{\alpha_{i_{l}}\in I}\alpha_{i_{l}}^{2}+\sum\limits_{\alpha_{i_{l}},\alpha_{i_{j}}\in I,i_{l}<i_{j}}\alpha_{i_{l}}\alpha_{i_{j}}\right)=( a_{3s}a_{1t}-a_{1s}a_{3t})\sum\limits_{\alpha_{i_{l}}\in I}\alpha_{i_{l}}+a_{2s}a_{3t}-a_{3s}a_{2t}(1\leq t<s\leq 3).$$
	\end{theorem}
	\begin{proof}
		By Lemma~\ref{monomiallyequivalent}, it suffices to consider $\mathrm{GRL}_k(\boldsymbol\alpha,\boldsymbol 1,
		\boldsymbol A_{3\times3})$. Write
		$$\boldsymbol{u}_{1}=\left(0,\ldots,0,a_{11},a_{21},a_{31}\right)^{T},\boldsymbol{u}_{2}=\left(0,\ldots,0,a_{12},a_{22},a_{32}\right)^{T}, \boldsymbol{u}_{3}=\left(0,\ldots,0,a_{13},a_{23},a_{33}\right)^{T}.$$ Since $\boldsymbol{G}_{4}$ is a parity-check matrix of $\mathrm{GRL}_{k}^{\perp}(\boldsymbol{\alpha},\boldsymbol{v},\boldsymbol{A}_{3\times 3})$, the code $\mathrm{GRL}_{k}^{\perp}(\boldsymbol{\alpha},\boldsymbol{v},\boldsymbol{A}_{3\times 3})$ is AMDS if  and only if it has parameters $\left[n+3,n+3-k,k\right]$. The characterization of minimum distance using a parity-check matrix shows that the minimum distance $d$ of  $\mathrm{GRL}_{k}^{\perp}(\boldsymbol{\alpha},\boldsymbol{v},\boldsymbol{A}_{3\times 3})$ equals $k$ if and only if the following two statements hold simultaneously:
		
		(I) any $k-1$ columns of $\boldsymbol{G}_{4}$ are $\mathbb{F}_{q}$-linearly independent;
		
		(II) there exist $k$ columns of $\boldsymbol{G}_{4}$ that are $\mathbb{F}_{q}$-linearly dependent.
		
		To establish (I), we consider the following six cases.
		
		\textbf{Case 1}. Suppose that the submatrix $\boldsymbol{M_{1}}$ formed by $k-1$ columns of $\boldsymbol{G}_{4}$ does not contain any of  $\boldsymbol{u}_{s}(1\leq s\leq 3)$, i.e., $$\boldsymbol{M_{1}}=\left(\begin{matrix}
			1&		1&		\cdots&		1&		1\\
			\alpha _{i_1}&		\alpha _{i_2}&		\cdots&		\alpha _{i_{k-2}}&		\alpha _{i_{k-1}}\\
			\vdots&		\vdots&		\quad&		\vdots&		\vdots\\
			\alpha _{i_1}^{k-4}&		\alpha _{i_2}^{k-4}&		\cdots&		\alpha _{i_{k-2}}^{k-4}&		\alpha _{i_{k-1}}^{k-4}\\
			\alpha _{i_1}^{k-3}&		\alpha _{i_2}^{k-3}&		\cdots&		\alpha _{i_{k-2}}^{k-3}&		\alpha _{i_{k-1}}^{k-3}\\
			\alpha _{i_1}^{k-2}&		\alpha _{i_2}^{k-2}&		\cdots&		\alpha _{i_{k-2}}^{k-2}&		\alpha _{i_{k-1}}^{k-2}\\
			\alpha _{i_1}^{k-1}&		\alpha _{i_2}^{k-1}&		\cdots&		\alpha _{i_{k-2}}^{k-1}&		\alpha _{i_{k-1}}^{k-1}\\
		\end{matrix} \right)_{k\times (k-1)}.$$
		Deleting the last row of $\boldsymbol{M_{1}}$ gives a nonsingular Vandermonde matrix. Hence the $k-1$ columns of $\boldsymbol{M_{1}}$ are $\mathbb{F}_{q}$-linearly independent.
		
		\textbf{Case 2}. Suppose that the submatrix $\boldsymbol{M_{2}}$ formed by $k-1$ columns of $\boldsymbol{G}_{4}$ contains exactly one of    $\boldsymbol{u}_{s}(1\leq s\leq 3)$, i.e., 
		$$\boldsymbol{M_{2}}=\left(\begin{matrix}
			1&		1&		\cdots&		1&		0\\
			\alpha _{i_1}&		\alpha _{i_2}&		\cdots&		\alpha _{i_{k-2}}&0\\
			\vdots&		\vdots&		\quad&		\vdots&		\vdots\\
			\alpha _{i_1}^{k-4}&		\alpha _{i_2}^{k-4}&		\cdots&		\alpha _{i_{k-2}}^{k-4}&0\\
			\alpha _{i_1}^{k-3}&		\alpha _{i_2}^{k-3}&		\cdots&		\alpha _{i_{k-2}}^{k-3}&		a_{1r}\\
			\alpha _{i_1}^{k-2}&		\alpha _{i_2}^{k-2}&		\cdots&		\alpha _{i_{k-2}}^{k-2}&a_{2r}\\
			\alpha _{i_1}^{k-1}&		\alpha _{i_2}^{k-1}&		\cdots&		\alpha _{i_{k-2}}^{k-1}&a_{3r}\\
		\end{matrix} \right)_{k\times (k-1)}.$$
		The $k-1$ columns of $\boldsymbol{M_{2}}$ are $\mathbb{F}_{q}$-linearly
		independent if and only if there exists a nonzero $(k-1)\times (k-1)$ minor of $\boldsymbol{M_{2}}$. We examine the following three submatrices.
		
		First, consider the matrix $\boldsymbol{R}_1$ obtained by deleting the last row of  $\boldsymbol{M_{2}}$, i.e., 
		$$\boldsymbol{R}_1=\left( \begin{matrix}
			1&		1&		\cdots&		1&		0\\
			\alpha _{i_1}&		\alpha _{i_2}&		\cdots&		\alpha _{i_{k-2}}&0\\
			\vdots&		\vdots&		\quad&		\vdots&		\vdots\\
			\alpha _{i_1}^{k-4}&		\alpha _{i_2}^{k-4}&		\cdots&		\alpha _{i_{k-2}}^{k-4}&0\\
			\alpha _{i_1}^{k-3}&		\alpha _{i_2}^{k-3}&		\cdots&		\alpha _{i_{k-2}}^{k-3}&		a_{1r}\\
			\alpha _{i_1}^{k-2}&		\alpha _{i_2}^{k-2}&		\cdots&		\alpha _{i_{k-2}}^{k-2}&a_{2r}\\
		\end{matrix} \right),$$
		Its determinant is 
		$$\mathrm{det}(\boldsymbol{R}_1)=\left(-a_{1r}\sum\limits_{\alpha _{i_l}\in I}\alpha _{i_l}+a_{2r}\right)\prod\limits_{1 \leq j <l\leq k-2}\left(\alpha _{i_l}-\alpha _{i_j}\right).$$
		Note that $\prod\limits_{1 \leq j< l\leq k-2}\left(\alpha_{i_{l}}-\alpha_{i_{j}}\right)\neq 0,$
		so $\mathrm{det}(\boldsymbol{R}_1)$ is a $(k-1)\times (k-1)$ nonzero minor of $\boldsymbol{M_{2}}$ if and only if for any subset $I\subseteq\left\{\alpha _{1},\ldots,\alpha _{n}\right\}$ of size $k-2$ and any $1\leq r\leq 3$, 
		$$-a_{1r}\sum\limits_{\alpha _{i_l}\in I}\alpha _{i_l}+a_{2r}\neq 0,$$
		i.e., 
		\begin{equation}\label{AMDScase21}
			a_{2r}\neq a_{1r}\sum\limits_{\alpha _{i_l}\in I}\alpha _{i_l}.
		\end{equation}
		
		Next, consider the matrix $\boldsymbol{R}_2$ obtained by deleting the $(k-1)$-th row of  $\boldsymbol{M_{2}}$, i.e., 
		$$\boldsymbol{R}_{2}=\left( \begin{matrix}
			1&		1&		\cdots&		1&		0\\
			\alpha _{i_1}&		\alpha _{i_2}&		\cdots&		\alpha _{i_{k-2}}&0\\
			\vdots&		\vdots&		\quad&		\vdots&		\vdots\\
			\alpha _{i_1}^{k-4}&		\alpha _{i_2}^{k-4}&		\cdots&		\alpha _{i_{k-2}}^{k-4}&0\\
			\alpha _{i_1}^{k-3}&		\alpha _{i_2}^{k-3}&		\cdots&		\alpha _{i_{k-2}}^{k-3}&		a_{1r}\\
			\alpha _{i_1}^{k-1}&		\alpha _{i_2}^{k-1}&		\cdots&		\alpha _{i_{k-2}}^{k-1}&a_{3r}\\
		\end{matrix} \right).$$
		By Lemma~\ref{lemma1},
		$$\mathrm{det}(\boldsymbol{R}_{2})=\left(-a_{1r}\left(\sum\limits_{\alpha_{i_{l}}\in I}\alpha_{i_{l}}^{2}+\sum\limits_{\alpha_{i_{l}},\alpha_{i_{j}}\in I,i_{l}<i_{j}}\alpha_{i_{l}}\alpha_{i_{j}}\right)+a_{3r}\right)\prod\limits_{1 \leq j<l\leq k-2}\left(\alpha _{i_l}-\alpha _{i_j}\right).$$
		Note that $\prod\limits_{1 \leq j< l\leq k-2}\left(\alpha_{i_{l}}-\alpha_{i_{j}}\right)\neq 0,$
		so $\mathrm{det}(\boldsymbol{R}_2)$ is a $(k-1)\times (k-1)$ nonzero minor of $\boldsymbol{M_{2}}$  if and only if for any subset $I\subseteq\left\{\alpha _{1},\ldots,\alpha _{n}\right\}$ of size $k-2$ and any $1\leq r\leq 3$, 
		$$-a_{1r}\left(\sum\limits_{\alpha_{i_{l}}\in I}\alpha_{i_{l}}^{2}+\sum\limits_{\alpha_{i_{l}},\alpha_{i_{j}}\in I,i_{l}<i_{j}}\alpha_{i_{l}}\alpha_{i_{j}}\right)+a_{3r}\neq 0,$$
		i.e., 
		\begin{equation}\label{AMDScase22}
			a_{3r}\neq a_{1r}\left(\sum\limits_{\alpha_{i_{l}}\in I}\alpha_{i_{l}}^{2}+\sum\limits_{\alpha_{i_{l}},\alpha_{i_{j}}\in I,i_{l}<i_{j}}\alpha_{i_{l}}\alpha_{i_{j}}\right).
		\end{equation}
		
		Finally, consider the matrix $\boldsymbol{R}_3$ obtained by deleting the $(k-2)$-th row of  $\boldsymbol{M_{2}}$, i.e., 
		$$\boldsymbol{R}_{3}=\left(\begin{matrix}
			1&		1&		\cdots&		1&		0\\
			\alpha _{i_1}&		\alpha _{i_2}&		\cdots&		\alpha _{i_{k-2}}&0\\
			\vdots&		\vdots&		\quad&		\vdots&		\vdots\\
			\alpha _{i_1}^{k-4}&		\alpha _{i_2}^{k-4}&		\cdots&		\alpha _{i_{k-2}}^{k-4}&0\\
			\alpha _{i_1}^{k-2}&		\alpha _{i_2}^{k-2}&		\cdots&		\alpha _{i_{k-2}}^{k-2}&		a_{2r}\\
			\alpha _{i_1}^{k-1}&		\alpha _{i_2}^{k-1}&		\cdots&		\alpha _{i_{k-2}}^{k-1}&a_{3r}\\
		\end{matrix} \right).$$
		By Lemma~\ref{lemma1},
		$$\mathrm{det}(\boldsymbol{R}_{3})=\left(-a_{2r}\left(\sum\limits_{\alpha_{i_{l}}\in I}\alpha_{i_{l}}^{2}+\sum\limits_{\alpha_{i_{l}},\alpha_{i_{j}}\in I,i_{l}<i_{j}}\alpha_{i_{l}}\alpha_{i_{j}}\right)+a_{3r}\sum\limits_{\alpha _{i_l}\in I}\alpha _{i_l} \right)\prod\limits_{1 \leq j<l\leq k-2}\left(\alpha _{i_l}-\alpha _{i_j}\right),$$
		Note that $\prod\limits_{1 \leq j< l\leq k-2}\left(\alpha_{i_{l}}-\alpha_{i_{j}}\right)\neq 0,$
		so $\mathrm{det}(\boldsymbol{R}_3)$ is a $(k-1)\times (k-1)$ nonzero minor of $\boldsymbol{M_{2}}$  if and only if for any subset $I\subseteq\left\{\alpha _{1},\ldots,\alpha _{n}\right\}$ of size $k-2$ and any $1\leq r\leq 3$, 
		$$-a_{2r}\left(\sum\limits_{\alpha_{i_{l}}\in I}\alpha_{i_{l}}^{2}+\sum\limits_{\alpha_{i_{l}},\alpha_{i_{j}}\in I,i_{l}<i_{j}}\alpha_{i_{l}}\alpha_{i_{j}}\right)+a_{3r}\sum\limits_{\alpha _{i_l}\in I}\alpha _{i_l} \neq 0,$$
		i.e., 
		\begin{equation}\label{AMDScase23}
			a_{3r}\sum\limits_{\alpha _{i_l}\in I}\alpha _{i_l}\neq a_{2r}\left(\sum\limits_{\alpha_{i_{l}}\in I}\alpha_{i_{l}}^{2}+\sum\limits_{\alpha_{i_{l}},\alpha_{i_{j}}\in I,i_{l}<i_{j}}\alpha_{i_{l}}\alpha_{i_{j}}\right).
		\end{equation}
		
		Conditions (\ref{AMDScase21})--(\ref{AMDScase23}) give condition $(1)$ of Theorem~\ref{AMDSDUALRLCODES}.
		
		\textbf{Case 3}. Suppose that the submatrix $\boldsymbol{M_{3}}$ formed by $k-1$ columns of $\boldsymbol{G}_{4}$ contains both $\boldsymbol{u}_{1}$ and $\boldsymbol{u}_{2}$, i.e.,
		$$\boldsymbol{M_{3}}=\left( \begin{matrix}
			1&		1&		\cdots&		1&0&		0\\
			\alpha _{i_1}&		\alpha _{i_2}&		\cdots&
			\alpha _{i_{k-3}}&	0&0\\
			\vdots&		\vdots&		\quad&		\vdots&		\vdots&		\vdots\\
			\alpha _{i_1}^{k-4}&		\alpha _{i_2}^{k-4}&		\cdots&
			\alpha _{i_{k-3}}^{k-4}&0&0\\
			\alpha _{i_1}^{k-3}&		\alpha _{i_2}^{k-3}&		\cdots&
			\alpha _{i_{k-3}}^{k-3}&a_{11}&		a_{12}\\
			\alpha _{i_1}^{k-2}&		\alpha _{i_2}^{k-2}&		\cdots&
			\alpha _{i_{k-3}}^{k-2}&a_{21}&a_{22}\\
			\alpha _{i_1}^{k-1}&		\alpha _{i_2}^{k-1}&		\cdots&
			\alpha _{i_{k-3}}^{k-1}&	a_{31}&a_{32}\\
		\end{matrix} \right)_{k\times (k-1)}.$$
		
		The $k-1$ columns of $\boldsymbol{M_{3}}$ are $\mathbb{F}_{q}$-linearly independent if and only if there exists a nonzero $(k-1)\times (k-1)$ minor of $\boldsymbol{M_{3}}$. We examine the following three submatrices.
		
		First, consider the matrix $\boldsymbol{S}_1$ obtained by deleting the last row of  $\boldsymbol{M_{3}}$, i.e., $$\boldsymbol{S}_{1}=\left(\begin{matrix}
			1&		1&		\cdots&		1&0&		0\\
			\alpha _{i_1}&		\alpha _{i_2}&		\cdots&
			\alpha _{i_{k-3}}&	0&0\\
			\vdots&		\vdots&		\quad&		\vdots&		\vdots&		\vdots\\
			\alpha _{i_1}^{k-4}&		\alpha _{i_2}^{k-4}&		\cdots&
			\alpha _{i_{k-3}}^{k-4}&0&0\\
			\alpha _{i_1}^{k-3}&		\alpha _{i_2}^{k-3}&		\cdots&
			\alpha _{i_{k-3}}^{k-3}&a_{11}&		a_{12}\\
			\alpha _{i_1}^{k-2}&		\alpha _{i_2}^{k-2}&		\cdots&
			\alpha _{i_{k-3}}^{k-2}&a_{21}&a_{22}\\
		\end{matrix} \right),$$
		Its determinant is 
		$$\mathrm{det}(\boldsymbol{S}_{1})=\left(a_{11}a_{22}-a_{12}a_{21}\right)\prod\limits_{1 \leq j<l\leq k-3}\left(\alpha _{i_l}-\alpha _{i_j}\right).$$
		Note that $\prod\limits_{1 \leq j< l\leq k-3}\left(\alpha_{i_{l}}-\alpha_{i_{j}}\right)\neq 0,$
		so $\mathrm{det}(\boldsymbol{S}_1)$ is a $(k-1)\times (k-1)$ nonzero minor of $\boldsymbol{M_{3}}$ if and only if 
		\begin{equation}\label{AMDScase31}
			a_{11}a_{22}-a_{12}a_{21}\neq 0.
		\end{equation}

		Next, consider the matrix $\boldsymbol{S}_2$ obtained by deleting the $(k-1)$-th row of  $\boldsymbol{M_{3}}$, i.e., 
		$$\boldsymbol{S}_{2}=\left(\begin{matrix}
			1&		1&		\cdots&		1&0&		0\\
			\alpha _{i_1}&		\alpha _{i_2}&		\cdots& \alpha _{i_{k-3}}&	0&0\\
			\vdots&		\vdots&		\quad&		\vdots&		\vdots&		\vdots\\
			\alpha _{i_1}^{k-4}&		\alpha _{i_2}^{k-4}&		\cdots&\alpha _{i_{k-3}}^{k-4}&0&0\\
			\alpha _{i_1}^{k-3}&		\alpha _{i_2}^{k-3}&		\cdots&\alpha _{i_{k-3}}^{k-3}&a_{11}&		a_{12}\\
			\alpha _{i_1}^{k-1}&		\alpha _{i_2}^{k-1}&		\cdots&\alpha _{i_{k-3}}^{k-1}&a_{31}&a_{32}\\
		\end{matrix} \right),$$
		Its determinant is $$\mathrm{det}(\boldsymbol{S}_{2})=\left(a_{11}a_{32}-a_{12}a_{31}\right)\prod\limits_{1 \leq j<l\leq k-3}\left(\alpha _{i_l}-\alpha _{i_j}\right).$$
		Note that $\prod\limits_{1 \leq j< l\leq k-3}\left(\alpha_{i_{l}}-\alpha_{i_{j}}\right)\neq 0,$ so $\mathrm{det}(\boldsymbol{S}_2)$ is a $(k-1)\times (k-1)$ nonzero minor of $\boldsymbol{M_{3}}$ if and only if 
		\begin{equation}\label{AMDScase32}
			a_{11}a_{32}-a_{12}a_{31}\neq 0.
		\end{equation}
		
		Finally, consider the matrix $\boldsymbol{S}_3$ obtained by deleting the $(k-2)$-th row of  $\boldsymbol{M_{3}}$, i.e., 
		$$\boldsymbol{S}_{3}=\left(\begin{matrix}1&		1&		\cdots&		1&0&		0\\
			\alpha _{i_1}&		\alpha _{i_2}&		\cdots&\alpha _{i_{k-3}}&	0&0\\
			\vdots&		\vdots&		\quad&		\vdots&		\vdots&		\vdots\\
			\alpha _{i_1}^{k-4}&		\alpha _{i_2}^{k-4}&		\cdots&\alpha _{i_{k-3}}^{k-4}&0&0\\
			\alpha _{i_1}^{k-2}&		\alpha _{i_2}^{k-2}&		\cdots&\alpha _{i_{k-3}}^{k-2}&a_{21}&a_{22}\\
			\alpha _{i_1}^{k-1}&		\alpha _{i_2}^{k-1}&		\cdots&\alpha _{i_{k-3}}^{k-1}&a_{31}&a_{32}\\
		\end{matrix} \right),$$
		Its determinant is 
		$$\mathrm{det}(\boldsymbol{S}_{3})=\left(a_{21}a_{32}-a_{22}a_{31}\right)\prod\limits_{1 \leq j<l\leq k-3}\left(\alpha _{i_l}-\alpha _{i_j}\right).$$
		Note that $\prod\limits_{1 \leq j< l\leq k-3}\left(\alpha_{i_{l}}-\alpha_{i_{j}}\right)\neq 0,$
		so $\mathrm{det}(\boldsymbol{S}_3)$ is a $(k-1)\times (k-1)$ nonzero minor of $\boldsymbol{M_{3}}$ if and only if 
		\begin{equation}\label{AMDScase33}
			a_{21}a_{32}-a_{22}a_{31}\neq 0.
		\end{equation}
		
		Since $\boldsymbol{A}_{3\times 3}=(a_{ij})_{3\times 3}\in\mathrm{GL}_{3}\left(\mathbb{F}_{q}\right)$, any two columns of $\boldsymbol{A}_{3\times 3}$ are $\mathbb{F}_q$-linearly independent. In particular, $\left(a_{11},a_{21},a_{31}\right)$ and $\left(a_{12},a_{22},a_{32}\right)$ are $\mathbb{F}_q$-linearly independent. Thus, there exists a nonzero $2\times2$ minor of $\begin{pmatrix}
			a_{11}&a_{12}\\
			a_{21}&a_{22}\\
			a_{31}&a_{32}
		\end{pmatrix}$. The three $2\times2$ minors of $\begin{pmatrix}
			a_{11}&a_{12}\\
			a_{21}&a_{22}\\
			a_{31}&a_{32}
		\end{pmatrix}$ are 
		$$\det\begin{pmatrix}
			a_{11}&a_{12}\\
			a_{21}&a_{22}
		\end{pmatrix}=a_{11}a_{22}-a_{12}a_{21}, \det\begin{pmatrix}
			a_{11}&a_{12}\\ 
			a_{31}&a_{32}
		\end{pmatrix}=a_{11}a_{32}-a_{12}a_{31}, \det\begin{pmatrix} 
			a_{21}&a_{22}\\
			a_{31}&a_{32}
		\end{pmatrix}=a_{21}a_{32}-a_{22}a_{31}.$$ 
		Hence, the $k-1$ columns of $\boldsymbol{M_{3}}$ are $\mathbb{F}_{q}$-linearly independent.
		
		\textbf{Case 4}. Suppose that the submatrix $\boldsymbol{M_{4}}$ formed by $k-1$ columns of $\boldsymbol{G}_{4}$ contains both $\boldsymbol{u}_{1}$ and $\boldsymbol{u}_{3}$, i.e.,
		$$\boldsymbol{M_{4}}=\left( \begin{matrix}
			1&		1&		\cdots&		1&0&		0\\
			\alpha _{i_1}&		\alpha _{i_2}&		\cdots&\alpha _{i_{k-3}}&	0&0\\
			\vdots&		\vdots&		\quad&		\vdots&		\vdots&		\vdots\\
			\alpha _{i_1}^{k-4}&		\alpha _{i_2}^{k-4}&		\cdots&\alpha _{i_{k-3}}^{k-4}&0&0\\
			\alpha _{i_1}^{k-3}&		\alpha _{i_2}^{k-3}&		\cdots&\alpha _{i_{k-3}}^{k-3}&a_{11}&		a_{13}\\
			\alpha _{i_1}^{k-2}&		\alpha _{i_2}^{k-2}&		\cdots&\alpha _{i_{k-3}}^{k-2}&a_{21}&a_{23}\\
			\alpha _{i_1}^{k-1}&		\alpha _{i_2}^{k-1}&		\cdots&\alpha _{i_{k-3}}^{k-1}&	a_{31}&a_{33}\\
		\end{matrix}  \right)_{k\times (k-1)}.$$
		The same argument as in Case $3$ shows that the $k-1$ columns of $\boldsymbol{M_{4}}$ are $\mathbb{F}_{q}$-linearly independent.
		
		\textbf{Case 5}. Suppose that the submatrix $\boldsymbol{M_{5}}$ formed by $k-1$ columns of $\boldsymbol{G}_{4}$ contains both $\boldsymbol{u}_{2}$ and $\boldsymbol{u}_{3}$, i.e.,
		$$\boldsymbol{M_{5}}=\left( \begin{matrix}
			1&		1&		\cdots&		1&0&		0\\
			\alpha _{i_1}&		\alpha _{i_2}&		\cdots&\alpha _{i_{k-3}}&	0&0\\
			\vdots&		\vdots&		\quad&		\vdots&		\vdots&		\vdots\\
			\alpha _{i_1}^{k-4}&		\alpha _{i_2}^{k-4}&		\cdots&
			\alpha _{i_{k-3}}^{k-4}&0&0\\
			\alpha _{i_1}^{k-3}&		\alpha _{i_2}^{k-3}&		\cdots&
			\alpha _{i_{k-3}}^{k-3}&a_{12}&		a_{13}\\
			\alpha _{i_1}^{k-2}&		\alpha _{i_2}^{k-2}&		\cdots&
			\alpha _{i_{k-3}}^{k-2}&a_{22}&a_{23}\\
			\alpha _{i_1}^{k-1}&		\alpha _{i_2}^{k-1}&		\cdots&
			\alpha _{i_{k-3}}^{k-1}&	a_{32}&a_{33}\\
		\end{matrix}\right)_{k\times (k-1)}.$$
		The same argument as in Case $3$ shows that the $k-1$ columns of $\boldsymbol{M_{5}}$ are $\mathbb{F}_{q}$-linearly independent.
		
		\textbf{Case 6}. Suppose that the submatrix $\boldsymbol{M_{6}}$ formed by $k-1$ columns of $\boldsymbol{G}_{4}$ contains all of $\boldsymbol{u}_{s}(1\leq s\leq 3)$, i.e.,
		$$\boldsymbol{M}_{6}=\left( \begin{matrix}
			1&		1&		\cdots&		1&0&0&		0\\
			\alpha _{i_1}&		\alpha _{i_2}&		\cdots&
			\alpha _{i_{k-4}}&0&	0&0\\
			\vdots&		\vdots&		\quad&		\vdots&		\vdots&		\vdots\\
			\alpha _{i_1}^{k-4}&		\alpha _{i_2}^{k-4}&		\cdots&
			\alpha _{i_{k-4}}^{k-4}&0&0&0\\
			\alpha _{i_1}^{k-3}&		\alpha _{i_2}^{k-3}&		\cdots&
			\alpha _{i_{k-4}}^{k-3}&a_{11}&		a_{12}&a_{13}\\
			\alpha _{i_1}^{k-2}&		\alpha _{i_2}^{k-2}&		\cdots&
			\alpha _{i_{k-4}}^{k-2}&a_{21}&a_{22}&a_{23}\\
			\alpha _{i_1}^{k-1}&		\alpha _{i_2}^{k-1}&		\cdots&
			\alpha _{i_{k-4}}^{k-1}&	a_{31}&a_{32}&a_{33}\\
		\end{matrix} \right)_{k\times (k-1)}.$$				Consider the matrix $\boldsymbol{S}_4$ obtained by deleting the $(k-3)$-th row of  $\boldsymbol{M_{6}}$, i.e.,
		$$\boldsymbol{S}_{4}=\left( \begin{matrix}
			1&		1&		\cdots&		1&0&0&		0\\
			\alpha _{i_1}&		\alpha _{i_2}&		\cdots&
			\alpha _{i_{k-4}}&0&	0&0\\
			\vdots&		\vdots&		\quad&		\vdots&		\vdots&		\vdots\\
			\alpha _{i_1}^{k-5}&		\alpha _{i_2}^{k-5}&		\cdots&
			\alpha _{i_{k-4}}^{k-5}&0&0&0\\
			\alpha _{i_1}^{k-3}&		\alpha _{i_2}^{k-3}&		\cdots&
			\alpha _{i_{k-4}}^{k-3}&a_{11}&		a_{12}&a_{13}\\
			\alpha _{i_1}^{k-2}&		\alpha _{i_2}^{k-2}&		\cdots&
			\alpha _{i_{k-4}}^{k-2}&a_{21}&a_{22}&a_{23}\\
			\alpha _{i_1}^{k-1}&		\alpha _{i_2}^{k-1}&		\cdots&
			\alpha _{i_{k-4}}^{k-1}&	a_{31}&a_{32}&a_{33}\\
		\end{matrix} \right)_{(k-1)\times (k-1)}.$$			
		Note that 
		$$\mathrm{det}(\boldsymbol{S}_{4})=\mathrm{det}(\boldsymbol{A}_{3\times 3})\prod\limits_{1 \leq j<l\leq k-4}\left(\alpha_{i_{l}}-\alpha_{i_{j}}\right)\neq 0,$$
		so the $k-1$ columns of $\boldsymbol{M}_{6}$ are $\mathbb{F}_{q}$-linearly independent.
		
		This establishes statement $(I).$
		
		By Lemma~\ref{MDSRLCODES}, statement $(II)$ is equivalent to condition $(2)$ of the theorem.
		
		This completes the proof. 
	\end{proof}
	\begin{corollary}\label{AMDSCODES2}
		Taking $\boldsymbol{A}_{3\times 3}=\left(\begin{matrix}
			\mu &\delta&1\\
			\tau&		1&0\\
			1&	0&0
		\end{matrix}\right)$ in Theorem $\ref{AMDSDUALRLCODES}$, where $\tau,\delta,\mu \in\mathbb{F}_{q}$, we obtain the following criterion: $\mathrm{GRL}_{k}^{\perp}(\boldsymbol{\alpha},\boldsymbol{v},\boldsymbol{A}_{3\times 3})$ is AMDS if and only if the following conditions hold simultaneously:
		
		$(1)$ for every subset $I\subseteq\left\{\alpha _{1},\ldots,\alpha _{n}\right\}$ of size $k-2$, the following conditions hold simultaneously:
		
		$(1.1)$ at least one of the following conditions holds:
		$$\tau\neq\mu \sum\limits_{\alpha _{i_l}\in I}\alpha _{i_l},\  1\neq\mu \left(\sum\limits_{\alpha_{i_{l}}\in I}\alpha_{i_{l}}^{2}+\sum\limits_{\alpha_{i_{l}},\alpha_{i_{j}}\in I,i_{l}<i_{j}}\alpha_{i_{l}}\alpha_{i_{j}}\right),\sum\limits_{\alpha _{i_l}\in I}\alpha _{i_l}\neq \tau\left(\sum\limits_{\alpha_{i_{l}}\in I}\alpha_{i_{l}}^{2}+\sum\limits_{\alpha_{i_{l}},\alpha_{i_{j}}\in I,i_{l}<i_{j}}\alpha_{i_{l}}\alpha_{i_{j}}\right);$$
		
		$(1.2)$ at least one of the following conditions holds:		$$\delta\sum\limits_{\alpha _{i_l}\in I}\alpha _{i_l}\neq 1,\  \delta\left(\sum\limits_{\alpha_{i_{l}}\in I}\alpha_{i_{l}}^{2}+\sum\limits_{\alpha_{i_{l}},\alpha_{i_{j}}\in I,i_{l}<i_{j}}\alpha_{i_{l}}\alpha_{i_{j}}\right)\neq 0,\ \sum\limits_{\alpha_{i_{l}}\in I}\alpha_{i_{l}}^{2}+\sum\limits_{\alpha_{i_{l}},\alpha_{i_{j}}\in I,i_{l}<i_{j}}\alpha_{i_{l}}\alpha_{i_{j}}\neq 0;$$
		
		$(1.3)$ at least one of the following conditions holds:		$$\sum\limits_{\alpha _{i_l}\in I}\alpha _{i_l}\neq 0,\  \sum\limits_{\alpha_{i_{l}}\in I}\alpha_{i_{l}}^{2}+\sum\limits_{\alpha_{i_{l}},\alpha_{i_{j}}\in I,i_{l}<i_{j}}\alpha_{i_{l}}\alpha_{i_{j}}\neq 0;$$
		
		$(2)$ at least one of the following conditions holds:
		
		$(2.1)$  there exists a subset $J\subseteq\left\{\alpha _{1},\ldots,\alpha _{n}\right\}$ of size $k-1$ such that at least one of the following conditions holds:
		$$\mu \sum\limits_{\alpha_{i_{l}},\alpha_{i_{j}}\in J,i_{l}<i_{j}}\alpha_{i_{l}}\alpha_{i_{j}}+1=\tau\sum\limits_{\alpha_{i_{l}}\in J}\alpha_{i_{l}},\ \delta\sum\limits_{\alpha_{i_{l}},\alpha_{i_{j}}\in J,i_{l}<i_{j}}\alpha_{i_{l}}\alpha_{i_{j}}=\sum\limits_{\alpha_{i_{l}}\in J}\alpha_{i_{l}},\ \sum\limits_{\alpha_{i_{l}},\alpha_{i_{j}}\in J,i_{l}<i_{j}}\alpha_{i_{l}}\alpha_{i_{j}}=0;$$
		
		$(2.2)$ there exists a subset $I\subseteq\left\{\alpha _{1},\ldots,\alpha _{n}\right\}$ of size $k-2$ such that at least one of the following conditions holds:
		$$\left(\mu -\tau\delta\right)\left(\sum\limits_{\alpha_{i_{l}}\in {I}}\alpha_{i_{l}}^{2}+\sum\limits_{\alpha_{i_{l}},\alpha_{i_{j}}\in I,i_{l}<i_{j}}\alpha_{i_{l}}\alpha_{i_{j}}\right)= -\delta\sum\limits_{\alpha_{i_{l}}\in {I}}\alpha_{i_{l}}+1,$$
		$$\tau\left(\sum\limits_{\alpha_{i_{l}}\in {I}}\alpha_{i_{l}}^{2}+\sum\limits_{\alpha_{i_{l}},\alpha_{i_{j}}\in I,i_{l}<i_{j}}\alpha_{i_{l}}\alpha_{i_{j}}\right)=\sum\limits_{\alpha_{i_{l}}\in {I}}\alpha_{i_{l}},$$
		$$\sum\limits_{\alpha_{i_{l}}\in {I}}\alpha_{i_{l}}^{2}+\sum\limits_{\alpha_{i_{l}},\alpha_{i_{j}}\in I,i_{l}<i_{j}}\alpha_{i_{l}}\alpha_{i_{j}}=0.$$ 
	\end{corollary} 								
	The following example illustrates Corollary~\ref{AMDSCODES2}.
	\begin{example}
		Let $(q,n,k)=(7,5,4), \boldsymbol{\alpha}=\left(1,2,3,4,5\right), \mu =2,\delta=4,\tau=3.$ For convenience, write  $$N=\sum\limits_{\alpha_{i_{l}}\in I}\alpha_{i_{l}}^2+\sum\limits_{\alpha_{i_{l}},\alpha_{i_{j}}\in I,i_{l}<i_{j}}\alpha_{i_{l}}\alpha_{i_{j}},$$
		A direct calculation gives the values in Table~\ref{table_example5.1}.
		
		\begin{table}[H]
			\centering 
			\caption{Verification of the conditions in Corollary~\ref{AMDSCODES2} over $\mathbb F_7$}
			\label{table_example5.1}
			\begin{tabular}{|c|c|c|c|c|c|c|c|}
				\hline	
				$I$&$\sum\limits_{\alpha_{i_{l}}\in I}\alpha_{i_{l}}$ &$\tau\neq\mu \sum\limits_{\alpha_{i_{l}}\in I}\alpha_{i_{l}}$&$\sum\limits_{\alpha_{i_{l}}\in I}\alpha_{i_{l}}^2$&$\sum\limits_{\alpha_{i_{l}},\alpha_{i_{j}}\in I,i_{l}<i_{j}}\alpha_{i_{l}}\alpha_{i_{j}}$&$N$&$1\neq \mu  N$&$\delta\sum\limits_{\alpha_{i_{l}}\in I}\alpha_{i_{l}}\neq 1$ \\
				\hline
				$\left\{1,2\right\}$&$3$&$3\neq 6$&$5$&$2$&$0$&&$5\neq 1$\\
				\hline
				$\left\{1,3\right\}$&$4$&$3\neq 1$&$3$&$3$&$6$& &$2\neq 1$\\
				\hline
				$\left\{1,4\right\}$&$5$&$3=3$&$3$ &$4$&$0$&$1\neq 0$&$ 6\neq 1$ \\
				\hline
				$\left\{1,5\right\}$&$6$&$3\neq 5$&$ 5$ &$5$&$3$&  &$3\neq 1$\\
				\hline
				$\left\{2,3\right\}$&$5$&$3=3$&$6$ &$6$&$5$&$1\neq 3$&$6\neq 1$\\
				\hline	
				$\left\{2,4\right\}$&$6$&$3\neq 5$&$ 6$ &$1$&$0$& &$3\neq 1$\\
				\hline	
				$\left\{2,5\right\}$&$0$&$3\neq 0$&$1$ &$3$&$4$& &$0\neq 1$\\
				\hline	
				$\left\{3,4\right\}$&$0$&$3\neq 0$&$4$ &$5$&$2$& &$0\neq 1$\\
				\hline	
				$\left\{3,5\right\}$&$1$&$3\neq 2$&$6$ &$1$&$0$& &$4 \neq 1$\\
				\hline
				$\left\{4,5\right\}$&$2$&$3\neq 4$&$6$ &$6$&$5$& &$1=1$\\
				\hline
			\end{tabular}
		\end{table}

		From Table \ref{table_example5.1}, when $I\notin\{\{1,4\},\{2,3\}\}$, we have $\tau\neq \mu\sum\limits_{\alpha_{i_l}\in I}\alpha_{i_l}$;
		when \(I=\{1,4\}\), we have $
		1\neq \mu N=0;$ when \(I=\{2,3\}\), we have $
		1\neq \mu N=10=3,$ so condition (1.1) of Corollary \ref{AMDSCODES2} holds for every subset \(I\subseteq
		\{\alpha_1,\ldots,\alpha_5\}\) of size \(k-2=2\); when \(I\neq\{4,5\}\), we have $\delta\sum\limits_{\alpha_{i_l}\in I}\alpha_{i_l}\neq 1$; when \(I=\{4,5\}\), we have $
		\delta N=20\neq 0.$ Thus, condition (1.2) of Corollary \ref{AMDSCODES2} also holds for every \(I\) of size $k-2=2$; when $I\notin\left\{\{2,5\},\{3,4\}\right\}$, we have $\sum\limits_{\alpha_{i_l}\in I}\alpha_{i_l}\neq 0$; when $I=\{2,5\}$, we have \(N=4\); when $I=\{3,4\},$ we have \(N=2\), so condition (1.3) of Corollary \ref{AMDSCODES2} holds for every \(I\). Moreover, the $3$-element subset $J=\{1,2,4\}\subseteq\{1,2,3,4,5\}$ satisfies 
		$\sum\limits_{\substack{\alpha_{i_l},\alpha_{i_j}\in J\\
				i_l<i_j}}
		\alpha_{i_l}\alpha_{i_j}
		=
		1\cdot2+1\cdot4+2\cdot4
		=
		14
		=
		0$ in \(\mathbb{F}_7\), so condition (2) of Corollary \ref{AMDSCODES2} holds. Consequently, 
		$$\boldsymbol{G}_{4}=\left(\begin{matrix}
			1&		1&1&		1&		1&		0&		0&		0\\
			1&2&3&4&5&2&4&1\\
			1&4&2&2&4&3&1&0\\
			1&1&6&1&6&1&0&0\\
		\end{matrix} \right)_{4\times 8}$$
		is a parity-check matrix of $\mathrm{GRL}_{k}^{\perp}(\boldsymbol{\alpha},\boldsymbol{v},\boldsymbol{A}_{3\times 3})$. Magma computations further show that $\mathrm{GRL}_{k}^{\perp}(\boldsymbol{\alpha},\boldsymbol{v},\boldsymbol{A}_{3\times 3})$ is an $\mathbb{F}_{7}$-linear code with parameters $\left[8,4,4\right]$.
	\end{example}	 
	\section{A necessary and sufficient condition for $\mathrm{GRL}_{k}\left(\boldsymbol{\alpha},\boldsymbol{v},\boldsymbol{A}_{3\times 3}\right)$ to be self-dual}\label{sec 5}
	In this section, we study the GRL code $\mathrm{GRL}_{k}\left(\boldsymbol{\alpha},\boldsymbol{v},\boldsymbol{A}_{3\times 3}\right)$ in Definition~\ref{definition1} and establish a necessary and sufficient condition for $\mathrm{GRL}_{k}\left(\boldsymbol{\alpha},\boldsymbol{v},\boldsymbol{A}_{3\times 3}\right)$ to be self-dual.  
	
	We begin by recalling a parity-check matrix of $\mathrm{GRL}_{k}\left(\boldsymbol{\alpha},\boldsymbol{v},\boldsymbol{A}_{3\times 3}\right)$ as follows.
	\begin{lemma}\label{paritycheckmatrix}{\rm(\cite{LFW}, Theorem 4.1)}
		Let $\mathbb{F}_q$ be the finite field with $q$ elements, where $q$ is a prime power. Let $u_{i}=\prod\limits_{j=1, j \neq i}^{n}\left(\alpha_{i}-\alpha_{j}\right)^{-1}(1 \leq i \leq n)$, and $ \boldsymbol{v}=\left(v_1,\ldots,v_n\right)\in(\mathbb{F}_{q}^{*})^n$. Then $\mathrm{GRL}_{k}\left(\boldsymbol{\alpha},\boldsymbol{v},\boldsymbol{A}_{3\times 3}\right)$ has a parity-check matrix 
		\begin{equation}\label{parity-check martix1}
			\boldsymbol{H}_{5}=\left(\begin{matrix}
				\frac{u_{1}}{v_{1}} & \frac{u_{2}}{v_{2}} & \cdots & \frac{u_{n}}{v_{n}} & 0 & 0& 0 \\
				\frac{u_{1}}{v_{1}} \alpha_{1} & \frac{u_{2}}{v_{2}} \alpha_{2} & \cdots & \frac{u_{n}}{v_{n}} \alpha_{n} & 0 & 0& 0 \\
				\vdots & \vdots & & \vdots & \vdots & \vdots& \vdots \\
				\frac{u_{1}}{v_{1}} \alpha_{1}^{n-k-1} & \frac{u_{2}}{v_{2}} \alpha_{2}^{n-k-1} & \cdots & \frac{u_{n}}{v_{n}} \alpha_{n}^{n-k-1} & 0 & 0& 0 \\
				\frac{u_{1}}{v_{1}} \alpha_{1}^{n-k} & \frac{u_{2}}{v_{2}} \alpha_{2}^{n-k} & \cdots & \frac{u_{n}}{v_{n}} \alpha_{n}^{n-k} & b_{11} & b_{12}& b_{13} \\
				\frac{u_{1}}{v_{1}} \alpha_{1}^{n-k+1} & \frac{u_{2}}{v_{2}} \alpha_{2}^{n-k+1} & \cdots & \frac{u_{n}}{v_{n}} \alpha_{n}^{n-k+1} & b_{21} & b_{22}& b_{23} \\
				\frac{u_{1}}{v_{1}} \alpha_{1}^{n-k+2} & \frac{u_{2}}{v_{2}} \alpha_{2}^{n-k+2} & \cdots & \frac{u_{n}}{v_{n}} \alpha_{n}^{n-k+2} & b_{31} & b_{32}& b_{33}
			\end{matrix}\right),
		\end{equation}
		where
		$$\boldsymbol{B}_{3\times 3}=\begin{pmatrix}
			b_{11}&		b_{12}&b_{13}\\
			b_{21}&		b_{22}&b_{23}\\
			b_{31}&		b_{32}&b_{33}
		\end{pmatrix}=\begin{pmatrix}
			0&0&-1\\
			0&-1&-\sum\limits_{i=1}^{n}\alpha_{i}\\
			-1&-\sum\limits_{i=1}^{n}\alpha_{i}&-\left(\sum\limits_{i=1}^{n}\alpha_{i}^{2}+\sum\limits_{1\leq i<j\leq n}\alpha_{i}\alpha_{j}\right)
		\end{pmatrix}^{T}\left(\boldsymbol{A}_{3\times 3}^{T}\right)^{-1}.$$
	\end{lemma} 
	
	The following lemma shows that equality of the full row spaces is equivalent to equality of the corresponding block row spaces. 
	\begin{lemma}\label{lem:block-separation}
		Let $n+3=2k$, and write \begin{equation}\label{GRLmatrix}
			\boldsymbol{G}_{5}=\left( \begin{matrix}
				v_1&		v_2&		\cdots&		v_{n-1}&		v_{n}&		0&		0&		0\\
				v_1\alpha _1&		v_2\alpha _2&		\cdots&		v_{n-1}\alpha _{n-1}&		v_{n}\alpha _n&		0&		0&		0\\
				\vdots&		\vdots&		\quad&		\vdots&		\vdots&		\vdots&		\vdots&		\vdots\\
				v_1\alpha _{1}^{k-4}&		v_2\alpha _{2}^{k-4}&		\cdots&		v_{n-1}\alpha _{n-1}^{k-4}&		v_{n}\alpha _{n}^{k-4}&		0&		0&		0\\
				v_1\alpha _{1}^{k-3}&		v_2\alpha _{2}^{k-3}&		\cdots&		v_{n-1}\alpha _{n-1}^{k-3}&		v_{n}\alpha _{n}^{k-3}&		a_{11}&		a_{12}&		a_{13}\\
				v_1\alpha _{1}^{k-2}&		v_2\alpha _{2}^{k-2}&		\cdots&		v_{n-1}\alpha _{n-1}^{k-2}&		v_{n}\alpha _{n}^{k-2}&		a_{21}&		a_{22}&		a_{23}\\
				v_1\alpha _{1}^{k-1}&		v_2\alpha _{2}^{k-1}&		\cdots&		v_{n-1}\alpha _{n-1}^{k-1}&		v_{n}\alpha _{n}^{k-1}&		a_{31}&		a_{32}&		a_{33}\\
			\end{matrix}\right)=\begin{pmatrix}
				\boldsymbol{g_{0}}\\
				\vdots\\
				\boldsymbol{g_{k-4}}\\
				\boldsymbol{g_{k-3}}\\ \boldsymbol{g_{k-2}}\\
				\boldsymbol{g_{k-1}}
			\end{pmatrix},
		\end{equation}
		and let $\boldsymbol{h_{0}},\ldots,\boldsymbol{h_{k-4}},\boldsymbol{h_{k-3}}, \boldsymbol{h_{k-2}},\boldsymbol{h_{k-1}}$ be the rows of $\boldsymbol H_5$. Then $$\mathrm{GRL}_{k}\left(\boldsymbol{\alpha},\boldsymbol{v},\boldsymbol{A}_{3\times 3}\right)=\operatorname{Span}_{\mathbb{F}_q}\left\{\boldsymbol{g_{0}},\ldots,\boldsymbol{g_{k-4}},\boldsymbol{g_{k-3}}, \boldsymbol{g_{k-2}},\boldsymbol{g_{k-1}}\right\}
		=\operatorname{Span}_{\mathbb{F}_q}\left\{\boldsymbol{h_{0}},\ldots,\boldsymbol{h_{k-4}},\boldsymbol{h_{k-3}}, \boldsymbol{h_{k-2}},\boldsymbol{h_{k-1}}\right\}$$
		if and only if 
		$$\operatorname{Span}_{\mathbb{F}_q}\left\{\boldsymbol{g_{0}},\ldots,\boldsymbol{g_{k-4}}\right\}
		=\operatorname{Span}_{\mathbb{F}_q}\left\{\boldsymbol{h_{0}},\ldots,\boldsymbol{h_{k-4}}\right\}$$
		and
		$$\operatorname{Span}_{\mathbb{F}_q}\left\{\boldsymbol{g_{k-3}}, \boldsymbol{g_{k-2}},\boldsymbol{g_{k-1}}\right\}
		=\operatorname{Span}_{\mathbb{F}_q}\left\{\boldsymbol{h_{k-3}}, \boldsymbol{h_{k-2}},\boldsymbol{h_{k-1}}\right\}$$
	\end{lemma}
	
	\begin{proof}
		Sufficiency follows by taking the sum of the two row spaces. It remains to prove necessity. 
		
		Suppose that $$\operatorname{Span}_{\mathbb{F}_q}\left\{\boldsymbol{g_{0}},\ldots,\boldsymbol{g_{k-4}},\boldsymbol{g_{k-3}}, \boldsymbol{g_{k-2}},\boldsymbol{g_{k-1}}\right\}
		=\operatorname{Span}_{\mathbb{F}_q}\left\{\boldsymbol{h_{0}},\ldots,\boldsymbol{h_{k-4}},\boldsymbol{h_{k-3}}, \boldsymbol{h_{k-2}},\boldsymbol{h_{k-1}}\right\},$$
		Then there exist $\tau_{0,i},\ldots,\tau_{k-1,i}\in\mathbb{F}_q$ and $\mu_{0,i},\ldots,\mu_{k-1,i}\in\mathbb{F}_q$ for $0\leq i\leq k-4$ such that 
		$$\boldsymbol{g}_{i}=\sum\limits_{s=0}^{k-4}\tau_{s,i}\boldsymbol{h}_{s}+\tau_{k-3,i}\boldsymbol{h}_{k-3}+\tau_{k-2,i}\boldsymbol{h}_{k-2}+\tau_{k-1,i}\boldsymbol{h}_{k-1}, 0\leq i\leq k-4,$$
		and
		$$\boldsymbol{h}_{i}=\sum\limits_{s=0}^{k-4}\mu_{s,i}\boldsymbol{g}_{s}+\mu_{k-3,i}\boldsymbol{g}_{k-3}+\mu_{k-2,i}\boldsymbol{g}_{k-2}+\mu_{k-1,i}\boldsymbol{g}_{k-1}, 0\leq i\leq k-4.$$
		Note that 
		$$\boldsymbol{g}_{i}=\begin{cases}
			\left( v_1\alpha _{1}^{i},v_2\alpha _{2}^{i},\ldots,v_{n}\alpha _{n}^{i},0,0,0\right),&0\leq i\leq k-4;\\
			\left( v_1\alpha _{1}^{k-4+s},v_2\alpha _{2}^{k-4+s},\ldots,v_{n}\alpha _{n}^{k-4+s},a_{s1},a_{s2},a_{s3}\right),&i=k-4+s,s=1,2,3,\\
		\end{cases}$$
		and 
		$$\boldsymbol{h}_{i}=\begin{cases}
			\left( \frac{u_{1}}{v_{1}}\alpha _{1}^{i},\frac{u_{2}}{v_{2}}\alpha _{2}^{i},\ldots,\frac{u_{n}}{v_{n}}\alpha _{n}^{i},0,0,0\right), &0\leq i\leq k-4;\\
			\left( \frac{u_{1}}{v_{1}}\alpha _{1}^{i},\frac{u_{2}}{v_{2}}\alpha _{2}^{i},\ldots,\frac{u_{n}}{v_{n}}\alpha _{n}^{i},b_{s1},b_{s2},b_{s3}\right), &i=k-4+s,s=1,2,3,
		\end{cases}$$
		Comparing the last three coordinates gives $$\begin{cases}
			\tau_{k-3,i}b_{11}+\tau_{k-2,i}b_{21}+\tau_{k-1,i}b_{31}=0\\
			\tau_{k-3,i}b_{12}+\tau_{k-2,i}b_{22}+\tau_{k-1,i}b_{32}=0\\
			\tau_{k-3,i}b_{13}+\tau_{k-2,i}b_{23}+\tau_{k-1,i}b_{33}=0
		\end{cases},$$
		and
		$$\begin{cases}
			\mu_{k-3,i}a_{11}+\mu_{k-2,i}a_{21}+\mu_{k-1,i}a_{31}=0\\
			\mu_{k-3,i}a_{12}+\mu_{k-2,i}a_{22}+\mu_{k-1,i}a_{32}=0\\
			\mu_{k-3,i}a_{13}+\mu_{k-2,i}a_{23}+\mu_{k-1,i}a_{33}=0
		\end{cases},$$
		i.e., $\begin{pmatrix}
			b_{11}&		b_{12}&b_{13}\\
			b_{21}&		b_{22}&b_{23}\\
			b_{31}&		b_{32}&b_{33}
		\end{pmatrix}^T\begin{pmatrix}
			\tau_{k-3,i}\\
			\tau_{k-2,i}\\
			\tau_{k-1,i}
		\end{pmatrix}=0$ and $\begin{pmatrix}
			a_{11}&		a_{12}&a_{13}\\
			a_{21}&		a_{22}&a_{23}\\
			a_{31}&		a_{32}&a_{33}
		\end{pmatrix}^T\begin{pmatrix}
			\mu_{k-3,i}\\
			\mu_{k-2,i}\\
			\mu_{k-1,i}
		\end{pmatrix}=0.$ Since $\boldsymbol A_{3\times3}\in\mathrm{GL}_{3}\left(\mathbb{F}_{q}\right)$, we also have $\boldsymbol B_{3\times3}\in\mathrm{GL}_{3}\left(\mathbb{F}_{q}\right)$. Consequently, $\tau_{k-3,i}=\tau_{k-2,i}=\tau_{k-1,i}=\mu_{k-3,i}=\mu_{k-2,i}=\mu_{k-1,i}=0$. It follows that the two sets of vectors $\boldsymbol{g_{0}},\ldots,\boldsymbol{g_{k-4}}$ and $\boldsymbol{h_{0}},\ldots,\boldsymbol{h_{k-4}}$ span the same $\mathbb{F}_q$-subspace, i.e., $\operatorname{Span}_{\mathbb{F}_q}\left\{\boldsymbol{g_{0}},\ldots,\boldsymbol{g_{k-4}}\right\}
		=\operatorname{Span}_{\mathbb{F}_q}\left\{\boldsymbol{h_{0}},\ldots,\boldsymbol{h_{k-4}}\right\}$.
		
		To prove the remaining equality $\operatorname{Span}_{\mathbb F_q}
		\left\{
		\boldsymbol g_{k-3},
		\boldsymbol g_{k-2},
		\boldsymbol g_{k-1}
		\right\}
		=
		\operatorname{Span}_{\mathbb F_q}
		\left\{
		\boldsymbol h_{k-3},
		\boldsymbol h_{k-2},
		\boldsymbol h_{k-1}
		\right\},$ we first establish the following claim.
		
		\textbf{Claim
			B} The two sets of vectors $\boldsymbol{g_{0}},\ldots,\boldsymbol{g_{k-4}}$ and $\boldsymbol{h_{0}},\ldots,\boldsymbol{h_{k-4}}$ span the same $\mathbb{F}_{q}$-subspace if and only if there exists a scalar $\lambda \in \mathbb{F}_{q}^{*}$ such that $v_{i}=\lambda\frac{u_{i}}{v_{i}}$ for every $1\leq i\leq n$. 
		
		If there exists a scalar $\lambda \in \mathbb{F}_{q}^{*}$ such that $v_{i}=\lambda\frac{u_{i}}{v_{i}}$  for every $1\leq i\leq n$, then
		$$\boldsymbol{h_{i}}=\lambda^{-1}\boldsymbol{g_{i}}(0\leq i\leq k-4), $$ 
		i.e., $\boldsymbol{g_{0}},\ldots,\boldsymbol{g_{k-4}}$ and $\boldsymbol{h_{0}},\ldots,\boldsymbol{h_{k-4}}$ span the same $\mathbb{F}_{q}$-subspace. Conversely, suppose that there exist $a_i$ and $b_i(0\leq i\leq k-4)$ such that $$\boldsymbol{g_{0}}=a_{0}\boldsymbol{h_{0}}+a_{1}\boldsymbol{h_{1}}+\cdots+a_{k-4}\boldsymbol{h_{k-4}}$$
		and
		$$\boldsymbol{g_{k-4}}=b_{0}\boldsymbol{h_{0}}+b_{1}\boldsymbol{h_{1}}+\cdots+b_{k-4}\boldsymbol{h_{k-4}},$$
		i.e.,
		$$v_{i}=\frac{u_{i}}{v_{i}}\left(a_{0}+a_{1}\alpha_{i}+\cdots+a_{k-4}\alpha_{i}^{k-4}\right)$$
		and
		$$v_{i}\alpha_{i}^{k-4}=\frac{u_{i}}{v_{i}}\left(b_{0}+b_{1}\alpha_{i}+\cdots+b_{k-4}\alpha_{i}^{k-4}\right).$$
		Define the polynomials $$f(x)=a_{0}+a_{1}x+\cdots+a_{k-4}x^{k-4}$$
		and 
		$$g(x)=b_{0}+b_{1}x+\cdots+b_{k-4}x^{k-4},$$
		Then
		$$f(\alpha_{i})=\frac{v_{i}^{2}}{u_{i}}(1\leq i\leq n)$$
		and
		$$\frac{v_{i}^{2}}{u_{i}}\alpha_{i}^{k-4}=g(\alpha_{i})(1\leq i\leq n),$$
		Thus,
		\begin{equation}\label{g=fa}
			g(\alpha_{i})=f(\alpha_{i})\alpha_{i}^{k-4}(1\leq i\leq n).
		\end{equation} 
		Set $r(x)=f(x)x^{k-4}-g(x)$. By $(\ref{g=fa})$,   $\alpha_1,\alpha_2,\ldots,\alpha_n$ are distinct roots of $r(x)$. Since $\deg(r(x))\leq 2k-8<n$, we obtain $r(x)=0$, i.e., $f(x)x^{k-4}=g(x)$. Comparing the coefficients of $f(x)x^{k-4}$ and $g(x)$ gives
		
		$$
		\left\{ \begin{array}{ll}
			a_0=b_{k-4},&\quad\\
			a_i=0,&\text{for}\ 1\leq i\leq k-4,\\
			b_j=0,&\text{for}\ 0\leq j\leq k-5.\\
		\end{array} \right.
		$$
		Hence, $f(x)=a_{0}$ and  $\boldsymbol{g_{0}}=a_{0}\boldsymbol{h_{0}}.$ Since $\boldsymbol{g_{0}}\neq \boldsymbol{0},$ we have $f(x)=a_{0}\in\mathbb{F}_{q}^{*}$. Consequently, for $1\leq i\leq n$, $$\frac{v_{i}^{2}}{u_{i}}=f(\alpha_{i})=a_{0}\in\mathbb{F}_{q}^{*},$$ 
		i.e., there exists a scalar $a_{0}\in\mathbb{F}_{q}^{*}$ such that $v_{i}=a_{0}\frac{u_{i}}{v_{i}}$ for every $1\leq i\leq n$, which proves Claim B.
		
		Now consider 
		$$\boldsymbol g_{k-4+s}=\left(v_1\alpha _{1}^{k-4+s},v_2\alpha _{2}^{k-4+s},\ldots,v_{n}\alpha _{n}^{k-4+s},	a_{s1},a_{s2},a_{s3}\right), s=1,2,3,$$
		and
		$$\boldsymbol h_{k-4+s}=\left(\frac{u_{1}}{v_{1}}\alpha _{1}^{k-4+s},\frac{u_{2}}{v_{2}}\alpha _{2}^{k-4+s},\ldots,	\frac{u_{n}}{v_{n}}\alpha _{n}^{k-4+s},	b_{s1},b_{s2},b_{s3}\right), s=1,2,3,$$
		Since there exists a scalar $\lambda \in \mathbb{F}_{q}^{*}$ such that $v_{i}=\lambda\frac{u_{i}}{v_{i}}$ for every $1\leq i\leq n$, the first $n$ coordinates of $\boldsymbol g_{k-4+s}-\lambda\boldsymbol h_{k-4+s}$ are all zero. Since $$\boldsymbol g_{k-4+s},\boldsymbol h_{k-4+s}\in 
		\mathrm{GRL}_{k}(\boldsymbol{\alpha}, \boldsymbol{v},\boldsymbol{A}_{l\times l}):=\left\{\left(v_{1} f\left(\alpha_{1}\right), \ldots, v_{n} f\left(\alpha_{n}\right),\boldsymbol{\beta}\right) | f(x) \in \mathbb{F}_{q}^{k}[x]\right\},
		$$
		their difference also belongs to the code: 
		$$\boldsymbol g_{k-4+s}-\lambda\boldsymbol h_{k-4+s}\in 
		\mathrm{GRL}_{k}(\boldsymbol{\alpha}, \boldsymbol{v},\boldsymbol{A}_{l\times l}):=\left\{\left(v_{1} f\left(\alpha_{1}\right), \ldots, v_{n} f\left(\alpha_{n}\right),\boldsymbol{\beta}\right) | f(x) \in \mathbb{F}_{q}^{k}[x]\right\},
		$$
		Thus, there exists a polynomial $f(x) \in \mathbb{F}_{q}^{k}[x]$ such that $f(\alpha_{i})=0(i=1,2,\ldots,n)$. It follows that $f(x)=0$, and hence $\boldsymbol g_{k-4+s}-\lambda\boldsymbol h_{k-4+s}=0$. Therefore,  
		\[
		\operatorname{Span}_{\mathbb F_q}
		\left\{
		\boldsymbol g_{k-3},
		\boldsymbol g_{k-2},
		\boldsymbol g_{k-1}
		\right\}
		=
		\operatorname{Span}_{\mathbb F_q}
		\left\{
		\boldsymbol h_{k-3},
		\boldsymbol h_{k-2},
		\boldsymbol h_{k-1}
		\right\}.
		\]
		
		This proves Lemma \ref{lem:block-separation}.
	\end{proof}
	\begin{theorem}\label{evenselfdualGRLcode}
		Let $\mathbb{F}_q$ be the finite field with $q$ elements, where $q$ is a prime power. Let $n+3=2k>6$,  $\boldsymbol{\alpha}=\left(\alpha_{1}, \ldots, \alpha_{n}\right) \in \mathbb{F}_{q}^{n}$ with $\alpha_{i} \neq \alpha_{j}(i \neq j), $ and $ u_{i}=\prod\limits_{j=1, j \neq i}^{n}\left(\alpha_{i}-\alpha_{j}\right)^{-1}(1 \leq i \leq n), \boldsymbol{v}=\left(v_1,\ldots,v_n\right)\in(\mathbb{F}_{q}^{*})^{n}$. Then $\mathrm{GRL}_{k}\left(\boldsymbol{\alpha},\boldsymbol{v},\boldsymbol{A}_{3\times 3}\right)$ is self-dual if and only if there exists a scalar $\lambda \in \mathbb{F}_{q}^{*}$ such that $v_{i}=\lambda\frac{u_{i}}{v_{i}}$ for every $1\leq i\leq n$, and
		\[
		\boldsymbol{A}_{3\times 3}\boldsymbol{A}_{3\times 3}^{T}=\lambda\left(\begin{matrix}
			0&0&-1\\
			0&-1&-\sum\limits_{i=1}^{n}\alpha_{i}\\
			-1&-\sum\limits_{i=1}^{n}\alpha_{i}&-\left(\sum\limits_{i=1}^{n}\alpha_{i}^{2}+\sum\limits_{1\leq i<j\leq n}\alpha_{i}\alpha_{j}\right)
		\end{matrix}\right).
		\] 
	\end{theorem}
	\begin{proof}
		By Definition~\ref{definition1}, $\mathrm{GRL}_{k}\left(\boldsymbol{\alpha},\boldsymbol{v},\boldsymbol{A}_{3\times 3}\right)$ has a generator matrix $\boldsymbol{G_{5}}$ given by $(\ref{GRLmatrix})$. By Lemma~\ref{paritycheckmatrix}, the matrix 
		\begin{equation}\label{parity-check martix2}
			\boldsymbol{H}_6=\left(\begin{matrix}
				\frac{u_{1}}{v_{1}} & \frac{u_{2}}{v_{2}} & \cdots & \frac{u_{n}}{v_{n}} & 0 & 0& 0 \\
				\frac{u_{1}}{v_{1}} \alpha_{1} & \frac{u_{2}}{v_{2}} \alpha_{2} & \cdots & \frac{u_{n}}{v_{n}} \alpha_{n} & 0 & 0& 0 \\
				\vdots & \vdots & & \vdots & \vdots & \vdots& \vdots \\
				\frac{u_{1}}{v_{1}} \alpha_{1}^{k-4} & \frac{u_{2}}{v_{2}} \alpha_{2}^{k-4} & \cdots & \frac{u_{n}}{v_{n}} \alpha_{n}^{k-4} & 0 & 0& 0 \\
				\frac{u_{1}}{v_{1}} \alpha_{1}^{k-3} & \frac{u_{2}}{v_{2}} \alpha_{2}^{k-3} & \cdots & \frac{u_{n}}{v_{n}} \alpha_{n}^{k-3} & b_{11} & b_{12}& b_{13} \\
				\frac{u_{1}}{v_{1}} \alpha_{1}^{k-2} & \frac{u_{2}}{v_{2}} \alpha_{2}^{k-2} & \cdots & \frac{u_{n}}{v_{n}} \alpha_{n}^{k-2} & b_{21} & b_{22}& b_{23} \\
				\frac{u_{1}}{v_{1}} \alpha_{1}^{k-1} & \frac{u_{2}}{v_{2}} \alpha_{2}^{k-1} & \cdots & \frac{u_{n}}{v_{n}} \alpha_{n}^{k-1} & b_{31} & b_{32}& b_{33}
			\end{matrix}\right)
		\end{equation}
		is a parity-check matrix of $\mathrm{GRL}_{k}\left(\boldsymbol{\alpha},\boldsymbol{v},\boldsymbol{A}_{3\times 3}\right)$, where
		$$\left(\begin{matrix}
			b_{11}&		b_{12}&b_{13}\\
			b_{21}&		b_{22}&b_{23}\\
			b_{31}&		b_{32}&b_{33}
		\end{matrix}\right)=\left(\begin{matrix}
			0&0&-1\\
			0&-1&-\sum\limits_{i=1}^{n}\alpha_{i}\\
			-1&-\sum\limits_{i=1}^{n}\alpha_{i}&-\left(\sum\limits_{i=1}^{n}\alpha_{i}^{2}+\sum\limits_{1\leq i<j\leq n}\alpha_{i}\alpha_{j}\right)
		\end{matrix}\right)^{T}\left(\boldsymbol{A}_{3\times 3}^{T}\right)^{-1}.$$
		
		Let $\boldsymbol{g_{i}}$ and $\boldsymbol{h_{i}}$ denote the $(i+1)$-th rows of $\boldsymbol{G}_{5}$ and  $\boldsymbol{H}_6$, respectively. By definition, $\mathrm{GRL}_{k}\left(\boldsymbol{\alpha},\boldsymbol{v},\boldsymbol{A}_{3\times 3}\right)$ is self-dual if and only if $\mathrm{GRL}_{k}\left(\boldsymbol{\alpha},\boldsymbol{v},\boldsymbol{A}_{3\times 3}\right)=\mathrm{GRL}_{k}^{\perp }\left(\boldsymbol{\alpha},\boldsymbol{v},\boldsymbol{A}_{3\times 3}\right)$. Equivalently,   $\boldsymbol{g_{0}},\ldots,\boldsymbol{g_{k-4}},\boldsymbol{g_{k-3}},\\ \boldsymbol{g_{k-2}},\boldsymbol{g_{k-1}}$ and $\boldsymbol{h_{0}},\ldots,\boldsymbol{h_{k-4}},\boldsymbol{h_{k-3}},$ $\boldsymbol{h_{k-2}},\boldsymbol{h_{k-1}}$ span the same $\mathbb{F}_{q}$-subspace. By Lemma \ref{lem:block-separation}, this is equivalent to the following two statements holding simultaneously.
		
		(i) $\boldsymbol{g_{0}},\ldots,\boldsymbol{g_{k-4}}$ and $ \boldsymbol{h_{0}},\ldots,\boldsymbol{h_{k-4}}$ span the same $\mathbb{F}_{q}$-subspace;
		
		(ii) $\boldsymbol{g_{k-3}},\boldsymbol{g_{k-2}},\boldsymbol{g_{k-1}}$ and $\boldsymbol{h_{k-3}},\boldsymbol{h_{k-2}},\boldsymbol{h_{k-1}}$ span the same $\mathbb{F}_{q}$-subspace. 
		
		By the proof of Lemma \ref{lem:block-separation}, statement (i) holds if and only if $v_{i}=\lambda\frac{u_{i}}{v_{i}}(\lambda\in\mathbb{F}_{q}^{*})$.
		
		Under this condition, statement (ii) holds if and only if   	$$\left(\begin{matrix}
			b_{11}&		b_{12}&b_{13}\\
			b_{21}&		b_{22}&b_{23}\\
			b_{31}&		b_{32}&b_{33}
		\end{matrix}\right)=\lambda^{-1}\left(\begin{matrix}
			a_{11}&		a_{12}&a_{13}\\
			a_{21}&		a_{22}&a_{23}\\
			a_{31}&		a_{32}&a_{33}
		\end{matrix}\right).
		$$
		Note that 
		$$\left(\begin{matrix}
			b_{11}&		b_{12}&b_{13}\\
			b_{21}&		b_{22}&b_{23}\\
			b_{31}&		b_{32}&b_{33}
		\end{matrix}\right)=\left(\begin{matrix}
			0&0&-1\\
			0&-1&-\sum\limits_{i=1}^{n}\alpha_{i}\\
			-1&-\sum\limits_{i=1}^{n}\alpha_{i}&-\left(\sum\limits_{i=1}^{n}\alpha_{i}^{2}+\sum\limits_{1\leq i<j\leq n}\alpha_{i}\alpha_{j}\right)
		\end{matrix}\right)\left(\left(\begin{matrix}
			a_{11}&		a_{12}&a_{13}\\
			a_{21}&		a_{22}&a_{23}\\
			a_{31}&		a_{32}&a_{33}
		\end{matrix}\right)^{T}\right)^{-1},$$
		Thus,  $\boldsymbol{g_{k-3}},\boldsymbol{g_{k-2}},\boldsymbol{g_{k-1}}$ and $\boldsymbol{h_{k-3}},\boldsymbol{h_{k-2}},\boldsymbol{h_{k-1}}$ span the same $\mathbb{F}_{q}$-subspace if and only if 
		$$\lambda^{-1}\left(\begin{matrix}
			a_{11}&		a_{12}&a_{13}\\
			a_{21}&		a_{22}&a_{23}\\
			a_{31}&		a_{32}&a_{33}
		\end{matrix}\right)=\left(\begin{matrix}
			0&0&-1\\
			0&-1&-\sum\limits_{i=1}^{n}\alpha_{i}\\
			-1&-\sum\limits_{i=1}^{n}\alpha_{i}&-\left(\sum\limits_{i=1}^{n}\alpha_{i}^{2}+\sum\limits_{1\leq i<j\leq n}\alpha_{i}\alpha_{j}\right)
		\end{matrix}\right)\left(\left(\begin{matrix}
			a_{11}&		a_{12}&a_{13}\\
			a_{21}&		a_{22}&a_{23}\\
			a_{31}&		a_{32}&a_{33}
		\end{matrix}\right)^{T}\right)^{-1},$$
		i.e.,	
		$$\left(\begin{matrix}
			a_{11}&		a_{12}&a_{13}\\
			a_{21}&		a_{22}&a_{23}\\
			a_{31}&		a_{32}&a_{33}
		\end{matrix}\right)\left(\begin{matrix}
			a_{11}&		a_{12}&a_{13}\\
			a_{21}&		a_{22}&a_{23}\\
			a_{31}&		a_{32}&a_{33}
		\end{matrix}\right)^{T}=\lambda\left(\begin{matrix}
			0&0&-1\\
			0&-1&-\sum\limits_{i=1}^{n}\alpha_{i}\\
			-1&-\sum\limits_{i=1}^{n}\alpha_{i}&-\left(\sum\limits_{i=1}^{n}\alpha_{i}^{2}+\sum\limits_{1\leq i<j\leq n}\alpha_{i}\alpha_{j}\right)
		\end{matrix}\right).$$
		
		This completes the proof. \end{proof}
	
	Examples $\ref{example4}$-$\ref{example5}$ illustrate Theorem $\ref{evenselfdualGRLcode}$ in the case $\boldsymbol{A}_{3\times 3}=\left(\begin{matrix}
		\mu &\tau&1\\
		\delta&		1&0\\
		1&	0&0
	\end{matrix}\right)$.
	\begin{example}\label{example4}
		Let $(q,n,k)=(13,5,4), \boldsymbol{\alpha}=\left(1,2,5,8,9\right),\mu =4,\delta=9$ and $\tau=3.$ A direct calculation gives $$\sum\limits_{i=1}^{5}\alpha_{i}=12,\sum\limits_{i=1}^{5}\alpha_{i}^{2}=6,\sum\limits_{1\leq j<i\leq 5}\alpha_{i}\alpha_{j}=4,$$ 
		$$ \boldsymbol{A}_{3\times 3}\boldsymbol{A}_{3\times 3}^{T}=\left(\begin{matrix}
			\mu ^2+\tau^2+1&\mu \delta+\tau&\mu \\
			\mu \delta+\tau&\delta^2+1&\delta\\
			\mu &\delta&1
		\end{matrix}\right)\\
		=\left(\begin{matrix}
			0&0&4\\
			0&4&9\\
			4&9&1
		\end{matrix}\right), 	$$
		and 
		$$\left(\begin{matrix}
			0&0&-1\\
			0&-1&-\sum\limits_{i=1}^{5}\alpha_{i}\\
			-1&-\sum\limits_{i=1}^{5}\alpha_{i}&-\left(\sum\limits_{i=1}^{5}\alpha_{i}^{2}+\sum\limits_{1\leq i<j\leq 5}\alpha_{i}\alpha_{j}\right)
		\end{matrix}\right)=\left(\begin{matrix}
			0&0&-1\\
			0&-1&1\\
			-1&1&-10
		\end{matrix}\right).$$
		These values satisfy 
		$$\boldsymbol{A}_{3\times 3}\boldsymbol{A}_{3\times 3}^{T}=\lambda\left(\begin{matrix}
			0&0&-1\\
			0&-1&-\sum\limits_{i=1}^{5}\alpha_{i}\\
			-1&-\sum\limits_{i=1}^{5}\alpha_{i}&-\left(\sum\limits_{i=1}^{5}\alpha_{i}^{2}+\sum\limits_{1\leq i<j\leq 5}\alpha_{i}\alpha_{j}\right)
		\end{matrix}\right)=\left(\begin{matrix}
			0&0&-9\\
			0&-9&9\\
			-9&9&1
		\end{matrix}\right)$$
		for $\lambda=9\in \mathbb{F}_{13}^{*}$.
		A further calculation shows that
		$$\boldsymbol{G}_5=\left(\begin{matrix}
			4&5&3&5&4&0&0&0\\
			4&10&2&1&10&4&3&1\\
			4&7&10&8&12&9&1&0\\
			4&1&11&12&4&1&0&0
		\end{matrix} \right)_{4\times 8}$$ 
		is a generator matrix of $\mathrm{GRL}_{k}\left(\boldsymbol{\alpha},\boldsymbol{v},\boldsymbol{A}_{3\times 3}\right)$. Furthermore, 
		Magma computations show that $\mathrm{GRL}_{k}\left(\boldsymbol{\alpha},\boldsymbol{v},\boldsymbol{A}_{3\times 3}\right)$ is an $\mathbb{F}_{13}$-linear code with parameters $\left[8,4,4\right]$. Thus, $\mathrm{GRL}_{k}\left(\boldsymbol{\alpha},\boldsymbol{v},\boldsymbol{A}_{3\times 3}\right)$ is an NMDS self-dual code over $\mathbb{F}_{13}$. 				\end{example}
	\begin{example}\label{example5}
		Let $(q,n,k)=(19,5,4), \boldsymbol{\alpha}=\left(0,1,2,16,18\right),\mu =12,\delta=7$ and $\tau=11.$ As in Example \ref*{example4}, we verify that there exists $\lambda=7\in \mathbb{F}_{19}^{*}$ such that $$\lambda\boldsymbol{u}=\lambda\left(u_1,u_2,u_3,u_4,u_5 \right)=
		(17,11,11,17,1)=\boldsymbol{v}^2=\left(v_{1}^{2},v_{2}^{2},v_{3}^{2},v_{4}^{2},v_{5}^{2}\right)=\left(6^{2},7^{2},7^{2},6^{2},1^{2}\right)$$ and  $$\boldsymbol{A}_{3\times 3}\boldsymbol{A}_{3\times 3}^{T}
		=\lambda\begin{pmatrix}
			0&0&-1\\
			0&-1&-\sum\limits_{i=1}^{5}\alpha_{i}\\
			-1&-\sum\limits_{i=1}^{5}\alpha_{i}&-\left(\sum\limits_{i=1}^{5}\alpha_{i}^{2}+\sum\limits_{1\leq i<j\leq 5}\alpha_{i}\alpha_{j}\right)
		\end{pmatrix}=\begin{pmatrix}
			0&0&12\\
			0&12&7\\
			12&7&1
		\end{pmatrix}.$$
		Consequently,
		$$\boldsymbol{G}_5=\left(\begin{matrix}
			6&7&7&6&1&0&0&0\\
			0&7&14&1&18&12&11&1\\
			0&7&9&16&1&7&1&0\\
			0&7&18&9&18&1&0&0
		\end{matrix} \right)_{4\times 8}$$ 
		is a generator matrix of $\mathrm{GRL}_{k}\left(\boldsymbol{\alpha},\boldsymbol{v},\boldsymbol{A}_{3\times 3}\right)$ . Furthermore, 
		Magma computations show that $\mathrm{GRL}_{k}\left(\boldsymbol{\alpha},\boldsymbol{v},\boldsymbol{A}_{3\times 3}\right)$ is an $\mathbb{F}_{19}$-linear code with parameters $\left[8,4,4\right]$, so $\mathrm{GRL}_{k}\left(\boldsymbol{\alpha},\boldsymbol{v},\boldsymbol{A}_{3\times 3}\right)$ is an NMDS self-dual code over $\mathbb{F}_{19}$. 
	\end{example}
	\begin{remark}
		The self-dual construction in \cite{LFW} requires \(q \equiv 1 (\bmod 4)\). Example \ref{example5}, however, gives a self-dual GRL code for $q\equiv 3(\bmod 4)$.   
	\end{remark} 
	
	\section{Comparison with several classes of non-GRS linear codes}\label{sec 6}
	
	In this section, we use Schur square dimensions to show that, under certain conditions, our GRL codes are not monomially equivalent to several classes of non-GRS linear codes with the same parameters $n,k,q$.
	Throughout this section, for $\mathrm{GRL}_k(\boldsymbol{\alpha},\boldsymbol{v},\boldsymbol{A}_{3\times3})$, we take
	$\boldsymbol{\alpha}=(\alpha_1,\ldots,\alpha_{n-3})$ with pairwise distinct entries in $\mathbb F_q$ and
	$\boldsymbol{v}=(v_1,\ldots,v_{n-3})\in(\mathbb F_q^*)^{n-3}$, so that the GRL code has length $n$.
	
	We first recall the definition of the Schur product.
	\begin{definition}\label{schurproduct}\textnormal{(\cite{ZhuTGRSFFA2024}, Definition 2.1)}
		For $\boldsymbol{x}=(x_1,\ldots,x_n)$ and $\boldsymbol{y}=(y_1,\ldots,y_n)$ in $\mathbb F_q^n$, their Schur product is defined as
		$$
		\boldsymbol{x}\star\boldsymbol{y}=(x_1y_1,\ldots,x_ny_n).
		$$
		The Schur product of two linear codes $\mathcal C_1,\mathcal C_2\subseteq\mathbb F_q^n$ is
		$$
		\mathcal C_1\star\mathcal C_2
		=\operatorname{Span}_{\mathbb F_q}
		\{\boldsymbol{c}_1\star\boldsymbol{c}_2:
		\boldsymbol{c}_1\in\mathcal C_1,\ \boldsymbol{c}_2\in\mathcal C_2\}.
		$$
		In particular, $\mathcal C\star\mathcal C$ is called the Schur square of $\mathcal C$ and is denoted by $\mathcal C^2$.
	\end{definition} 
	
	We next recall the definition of twisted generalized Reed--Solomon codes given by Hu et al.
	\begin{definition}\label{LPTGRSdefinition}\textnormal{(\cite{HuTGRSTIT2025}, Definition 2)}
		Let $n$ and $k$ be integers with $0<k\le n$, let $\boldsymbol{B}=(b_{i,j})\in\mathbb F_q^{k\times(n-k)}$, and let
		$\mathcal L\subseteq\{0,1,\ldots,n-k-1\}$ and
		$\mathcal P\subseteq\{0,1,\ldots,k-1\}$.
		Let $\boldsymbol{\beta}=(\beta_1,\ldots,\beta_n)\in\mathbb F_q^n$ have pairwise distinct entries, and let
		$\boldsymbol{u}=(u_1,\ldots,u_n)\in(\mathbb F_q^*)^n$.
		The $(\mathcal L,\mathcal P)$-twisted generalized Reed--Solomon code, abbreviated as an $(\mathcal L,\mathcal P)$-TGRS code, is defined as
		$$
		\mathcal C(\mathcal L,\mathcal P,\boldsymbol B)
		=\{(u_1f(\beta_1),\ldots,u_nf(\beta_n)):
		f(x)\in\mathcal F(\mathcal L,\mathcal P,\boldsymbol B)\},
		$$
		where
		$$
		\mathcal F(\mathcal L,\mathcal P,\boldsymbol B)
		=\left\{
		\sum_{i=0}^{k-1}f_ix^i
		+\sum_{i\in\mathcal P}f_i\sum_{j\in\mathcal L}b_{i,j}x^{k+j}:
		f_0,\ldots,f_{k-1}\in\mathbb F_q
		\right\}.
		$$
	\end{definition}
	
	We now establish a sufficient condition for the Schur square of a $[n,k]_q$ GRL code to have dimension $2k+2$.
	
	\begin{theorem}\label{thm:GRL-Schur-square}
		Let $k\ge4$ and $n\ge2k+2$, and define
		$$
		\boldsymbol{A}_{3\times3}^{\star2}
		=\begin{pmatrix}
			a_{11}^2&a_{12}^2&a_{13}^2\\
			a_{11}a_{21}&a_{12}a_{22}&a_{13}a_{23}\\
			a_{11}a_{31}-a_{21}^2&
			a_{12}a_{32}-a_{22}^2&
			a_{13}a_{33}-a_{23}^2
		\end{pmatrix}.
		$$
		If $\operatorname{rank}(\boldsymbol{A}_{3\times3}^{\star2})=3$, then
		$$
		\dim\left(\mathrm{GRL}_k^2(\boldsymbol{\alpha},\boldsymbol{v},\boldsymbol{A}_{3\times3})\right)=2k+2.
		$$
	\end{theorem}
	
	\begin{proof}
		Let $\boldsymbol{g}_0,\ldots,\boldsymbol{g}_{k-1}$ be the rows of the generator matrix of
		$\mathrm{GRL}_k(\boldsymbol{\alpha},\boldsymbol{v},\boldsymbol{A}_{3\times3})$.
		By Definition~\ref{schurproduct},
		$$
		\mathrm{GRL}_k^2(\boldsymbol{\alpha},\boldsymbol{v},\boldsymbol{A}_{3\times3})
		=\operatorname{Span}_{\mathbb F_q}
		\{\boldsymbol{g}_i\star\boldsymbol{g}_j:0\le i\le j\le k-1\}.
		$$
		A direct calculation gives
		$$
		\begin{aligned}
			&\mathrm{GRL}_k^2(\boldsymbol{\alpha},\boldsymbol{v},\boldsymbol{A}_{3\times3})\\
			={}&\operatorname{Span}_{\mathbb F_q}\left\{
			(\boldsymbol{w}_z,0,0,0)\ (0\le z\le2k-5),\;
			(\boldsymbol{w}_{2k-6},a_{11}^2,a_{12}^2,a_{13}^2),(\boldsymbol{w}_{2k-5},a_{11}a_{21},a_{12}a_{22},a_{13}a_{23}),\right.\\
			&\quad 
			(\boldsymbol{w}_{2k-4},a_{11}a_{31},a_{12}a_{32},a_{13}a_{33}), (\boldsymbol{w}_{2k-4},a_{21}^2,a_{22}^2,a_{23}^2),\;
			(\boldsymbol{w}_{2k-3},a_{21}a_{31},a_{22}a_{32},a_{23}a_{33}),\;
			\left.(\boldsymbol{w}_{2k-2},a_{31}^2,a_{32}^2,a_{33}^2)
			\right\},
		\end{aligned}
		$$
		where $\boldsymbol{w}_z=(v_1^2\alpha_1^z,\ldots,v_{n-3}^2\alpha_{n-3}^z)$ for every non-negative integer $z$. Thus,
		$$
		\begin{aligned}
			\dim\left(\mathrm{GRL}_k^2(\boldsymbol{\alpha},\boldsymbol{v},\boldsymbol{A}_{3\times3})\right)
			=&\operatorname{rank}\begin{pmatrix}
				\boldsymbol{w}_0&0&0&0\\
				\boldsymbol{w}_1&0&0&0\\
				\vdots&\vdots&\vdots&\vdots\\
				\boldsymbol{w}_{2k-5}&0&0&0\\
				\boldsymbol{w}_{2k-6}&a_{11}^2&a_{12}^2&a_{13}^2\\
				\boldsymbol{w}_{2k-5}&a_{11}a_{21}&a_{12}a_{22}&a_{13}a_{23}\\
				\boldsymbol{w}_{2k-4}&a_{11}a_{31}&a_{12}a_{32}&a_{13}a_{33}\\
				\boldsymbol{w}_{2k-4}&a_{21}^2&a_{22}^2&a_{23}^2\\
				\boldsymbol{w}_{2k-3}&a_{21}a_{31}&a_{22}a_{32}&a_{23}a_{33}\\
				\boldsymbol{w}_{2k-2}&a_{31}^2&a_{32}^2&a_{33}^2
			\end{pmatrix}\\
			=&\operatorname{rank}\begin{pmatrix}
				\boldsymbol{w}_0&0&0&0\\
				\boldsymbol{w}_1&0&0&0\\
				\vdots&\vdots&\vdots&\vdots\\
				\boldsymbol{w}_{2k-5}&0&0&0\\
				\boldsymbol{w}_{2k-4}&a_{21}^2&a_{22}^2&a_{23}^2\\
				\boldsymbol{w}_{2k-3}&a_{21}a_{31}&a_{22}a_{32}&a_{23}a_{33}\\
				\boldsymbol{w}_{2k-2}&a_{31}^2&a_{32}^2&a_{33}^2\\
				\boldsymbol{0}_{1\times(n-3)}&a_{11}^2&a_{12}^2&a_{13}^2\\
				\boldsymbol{0}_{1\times(n-3)}&a_{11}a_{21}&a_{12}a_{22}&a_{13}a_{23}\\
				\boldsymbol{0}_{1\times(n-3)}&a_{11}a_{31}-a_{21}^2&a_{12}a_{32}-a_{22}^2&a_{13}a_{33}-a_{23}^2
			\end{pmatrix}\\
			=&\operatorname{rank}\begin{pmatrix}
				\boldsymbol{T}_1&\boldsymbol{T}_2\\
				\boldsymbol{0}_{3\times(n-3)}&\boldsymbol{A}_{3\times3}^{\star2}
			\end{pmatrix},
		\end{aligned}
		$$
		where
		$$
		\boldsymbol{T}_1=\begin{pmatrix}
			v_1^2&\cdots&v_{n-3}^2\\
			v_1^2\alpha_1&\cdots&v_{n-3}^2\alpha_{n-3}\\
			\vdots&&\vdots\\
			v_1^2\alpha_1^{2k-2}&\cdots&v_{n-3}^2\alpha_{n-3}^{2k-2}
		\end{pmatrix},\qquad
		\boldsymbol{T}_2=\begin{pmatrix}
			\multicolumn{3}{c}{\boldsymbol{0}_{(2k-4)\times3}}\\
			a_{21}^2&a_{22}^2&a_{23}^2\\
			a_{21}a_{31}&a_{22}a_{32}&a_{23}a_{33}\\
			a_{31}^2&a_{32}^2&a_{33}^2
		\end{pmatrix}.
		$$
		Since $\operatorname{rank}(\boldsymbol{A}_{3\times3}^{\star2})=3$, the matrix $\boldsymbol{A}_{3\times3}^{\star2}$ is invertible. Elementary row operations therefore give
		$$
		\begin{aligned}
			&\dim\left(\mathrm{GRL}_k^2(\boldsymbol{\alpha},\boldsymbol{v},\boldsymbol{A}_{3\times3})\right)\\
			={}&\operatorname{rank}\left[
			\begin{pmatrix}
				\boldsymbol I_{2k-1}&-\boldsymbol T_2(\boldsymbol A_{3\times3}^{\star2})^{-1}\\
				\boldsymbol0_{3\times(2k-1)}&\boldsymbol I_3
			\end{pmatrix}
			\begin{pmatrix}
				\boldsymbol T_1&\boldsymbol T_2\\
				\boldsymbol0_{3\times(n-3)}&\boldsymbol A_{3\times3}^{\star2}
			\end{pmatrix}\right]\\
			={}&\operatorname{rank}\begin{pmatrix}
				\boldsymbol T_1&\boldsymbol0_{(2k-1)\times3}\\
				\boldsymbol0_{3\times(n-3)}&\boldsymbol A_{3\times3}^{\star2}
			\end{pmatrix}\\
			={}&\operatorname{rank}(\boldsymbol T_1)
			+\operatorname{rank}(\boldsymbol A_{3\times3}^{\star2}).
		\end{aligned}
		$$
		Since $n\geq 2k+2$, we have $n-3\ge2k-1$ so  $\mathrm{rank}\left(\boldsymbol T_1\right)=2k-1$.
		Therefore,
		$$
		\dim\left(\mathrm{GRL}_k^2(\boldsymbol{\alpha},\boldsymbol{v},\boldsymbol{A}_{3\times3})\right)
		=(2k-1)+3=2k+2.
		$$
	\end{proof}
	
	It is well known that equivalent linear codes have Schur squares of the same dimension.
	
	Table~\ref{Schur2TGRS} summarizes known results on Schur square dimensions of TGRS codes corresponding to several special choices of $\boldsymbol{B}$. 
	Table~\ref{Schur2NonGRS} lists known results on Schur square dimensions for several other classes of non-GRS linear codes of length $n$ and dimension $k$.   Together with Theorem~\ref{thm:GRL-Schur-square}, these results show that our GRL codes are not monomially equivalent to the listed code families in certain cases.
	
	\begin{table}[H]
		\centering
		\small
		\setlength{\tabcolsep}{3pt}
		\renewcommand{\arraystretch}{1}
		\setlength{\bigstrutjot}{16pt}
		\caption{Known Schur square dimensions of $(\mathcal L,\mathcal P)$-TGRS codes of length $n$ and dimension $k$}
		\label{Schur2TGRS}
		\begin{tabular}{|c|c|c|c|}
			\hline
			\rule[-4pt]{0pt}{18pt}
			$\boldsymbol B$&Range of $n,k$&\makecell{Dimension of\\the Schur square}&Reference\\
			\hline
			\bigstrut
			$\begin{pmatrix}
				b_{0,0}&\boldsymbol0_{1\times(n-k-1)}\\
				\boldsymbol0_{(k-1)\times1}&\boldsymbol0_{(k-1)\times(n-k-1)}
			\end{pmatrix}$
			&$3\le k<\dfrac n2$&$2k$&\cite{liu2021construction}\\
			\hline
			\bigstrut
			$\begin{pmatrix}
				\boldsymbol0_{(k-1)\times1}&\boldsymbol0_{(k-1)\times(n-k-1)}\\
				b_{k-1,0}&\boldsymbol0_{1\times(n-k-1)}
			\end{pmatrix}$
			&$3\le k<\dfrac{n+1}2$&$2k$&\cite{liu2021construction}\\
			\hline
			\bigstrut
			$\begin{pmatrix}
				\boldsymbol0_{(k-2)\times1}&\boldsymbol0_{(k-2)\times1}&\boldsymbol0_{(k-2)\times(n-k-2)}\\
				b_{k-2,0}&b_{k-2,1}&\boldsymbol0_{1\times(n-k-2)}\\
				b_{k-1,0}&b_{k-1,1}&\boldsymbol0_{1\times(n-k-2)}
			\end{pmatrix}$
			&$4\le k\le\dfrac{n-3}2$&$\in\left\{2k,2k+1,2k+2\right\}$& \cite{ZhuTGRSFFA2024}\\
			\hline
			\multirow{2}[tb4]{*}{
				$\begin{pmatrix}
					\boldsymbol0_{(k-l)\times l}&\boldsymbol0_{(k-l)\times(n-k-l)}\\
					\begin{matrix}
						b_{k-l,0}&0&\cdots&0\\
						b_{k-l+1,0}&b_{k-l+1,1}&\cdots&0\\
						\vdots&\vdots&\ddots&\vdots\\
						b_{k-1,0}&b_{k-1,1}&\cdots&b_{k-1,l-1}
					\end{matrix}&\boldsymbol0_{l\times(n-k-l)}
				\end{pmatrix}$}
			&\bigstrut $2l+2\le k\le\dfrac{n-2l+2}2$&$\geq 2k+2l-2$&\cite{HuTGRSTIT2025}\\
			\cline{2-4}
			&\bigstrut $k\le\min\left\{2l+2,\dfrac{n-2l+2}2\right\}$&$\geq 3k-3$&\cite{HuTGRSTIT2025}\\
			\hline
		\end{tabular}
	\end{table}

\begin{table}[H]
	\centering
	\small
	\setlength{\tabcolsep}{4pt}
	\renewcommand{\arraystretch}{2.3}
	\caption{Schur square dimensions of other non-GRS
		linear codes of length $n$ and dimension $k$}
	\label{Schur2NonGRS}
	\resizebox{\linewidth}{!}{%
		\begin{tabular}{|c|c|c|c|}
			\hline
			Code family
			& Conditions
			& \makecell{Dimension of\\the Schur square}
			& Reference\\
			\hline
			Roth--Lempel codes
			& $3\le k\le\dfrac n2$
			& $2k$
			& \cite[Theorem IV.3]{XuZhouRL2026}\\
			\hline
			One-point elliptic curve codes
			& $3\le k<\dfrac n2$
			& $2k$
			& \cite[Theorem 4]{ChenAG2024}\\
			\hline
			\makecell{One-point algebraic\\geometry codes of genus $g$}
			& $g\ge2,\ g\ne3,\ 4g<k+g-1<\dfrac n2$
			& $2k+g-1$
			& \cite[Theorem 6]{ChenAG2024}\\
			\hline
			Two-point elliptic curve codes
			& $4\le k<\dfrac n2,\ k\equiv0\pmod2$
			& $2k$
			& \cite[Theorem 3]{WangElliptic2026}\\
			\hline
			\multirow{2}{*}{Deleted GRS codes}
			& $3\le k<\dfrac n2,\ l=1,\ h\in\{1,k-1\}$
			& $2k$
			& \multirow{2}{*}{
				\cite[Lemma 10]{WanZhuDGRS2026}}\\
			\cline{2-3}
			& $4\le k<\dfrac{n-1}{2},\ l=1,\ 2\le h\le k-2$
			& $2k+1$
			&\\
			\hline
			\multirow{3}{*}{Linear codes $\mathcal C_{i,j}$}
			& $5\le k\le\dfrac{n-4}{2},\ i=1,\ j\in\{2,3,k\}$
			& $2k+1$
			& \multirow{3}{*}{
				\cite[Theorem 3.7]{AbdukhalikovVerma2026}}\\
			\cline{2-3}
			& $5\le k\le\dfrac{n-4}{2},\
			i\in\{k-2,k-1\},\ j=k$
			& $2k+1$
			&\\
			\cline{2-3}
			& $5\le k\le\dfrac{n-4}{2},\
			2\le i\le k-3,\ 4\le j\le k-1,\ i<j$
			& $2k+3$
			&\\
			\hline
			Linear codes $\mathcal C_k(S,\boldsymbol v,\infty)$
			& $3\le k<\dfrac{n-1}{2}$
			& $2k$
			& \cite[Theorem III.6(1)]{LiSunZhu2025}\\
			\hline
			Linear codes $\mathcal C_k(A,\boldsymbol v)$
			& $3\le k<\dfrac n2$
			& $2k$
			& \cite[Theorem 4.4]{A23}\\
			\hline
			\multirow{2}{*}{
				\makecell{Column-twisted\\Reed--Solomon codes}}
			& $3\le k\le\dfrac{n-2}{2}$
			& $2k$
			& \cite[Theorem 2]{LiuColumnTRS2026}\\
			\cline{2-4}
			& $3\le k\le\dfrac n2$
			& $2k$
			& \cite[Theorem 3.5(a)]{BhagatKaipaCTRS2026}\\
			\hline
			\makecell{Linear codes\\
				$\operatorname{CTRS}
				(\mathcal A,\boldsymbol B,\boldsymbol\lambda)$}
			& $4\le k\le\dfrac{n-1}{2}$
			& $2k+1$
			& \cite[Theorem 3.3(ii)]{GuLeiZhangVariants2026}\\
			\hline
			\makecell{Linear codes\\
				$\operatorname{RCTRS}_{l,\eta}
				(\mathcal A,\boldsymbol B,\boldsymbol\lambda)$}
			& $7\le k<\dfrac n2-1,\ 1\le l\le k-2$
			& $2k+3$
			& \cite[Theorem 4.4(ii)]{GuLeiZhangVariants2026}\\
			\hline
			\makecell{Linear codes\\
				$\operatorname{RCTRS}_{0,1}
				(\boldsymbol\alpha,b,c,\lambda,\eta)$}
			& $4\le k\le\dfrac{n-1}{2}$
			& $2k+1$
			& \cite[Theorem 4.10]{BhagatSinghSarmaRCTRS}\\
			\hline
			\makecell{Linear codes\\
				$\operatorname{RCTRS}_{k-1,1}
				(\boldsymbol\alpha,b,c,\lambda,\eta)$}
			& $4\le k\le\dfrac{n-1}{2},\
			\operatorname{char}(\mathbb F_q)\ne2$
			& $2k+1$
			& \cite[Theorem 5.3]{BhagatSinghSarmaRCTRS}\\
			\hline
			\multirow{2}{*}{Twisted Roth--Lempel codes}
			& $5\le k\le\dfrac{n-3}{2},\ l\in\{k-2,k-1\}$
			& $2k+1$
			& \multirow{2}{*}{
				\cite[Theorem 3.11]{LeiTRL2026}}\\
			\cline{2-3}
			& $5\le k\le\dfrac{n-3}{2},\ 1\le l\le k-3$
			& $2k+3$
			&\\
			\hline
			\makecell{Hyperderivative\\Reed--Solomon codes}
			& $n=rs,\ s\ge2,\ r\ge2k-2s+1,\
			p\ge k\ge\max\{5,2s\}$
			& $(2k-2s+1)s$
			& \cite[Theorem 3.2(ii)]{GuZhuZhangHRS2026}\\
			\hline
			Cyclic MDS codes
			& $n\ge7,\ k=|J|\ge3,\ 2\max J<n,\
			|J+J|\ne2k+2$
			& $|J+J|$
			& \cite[Theorem 11]{XiangTangCyclic2026}\\
			\hline
		\end{tabular}%
	}
	\par\smallskip
	\begin{minipage}{\linewidth}
		\footnotesize
		Note: All notation and additional hypotheses are as
		in the corresponding references.
	\end{minipage}
\end{table}
	
	\section{Conclusions}\label{sec 7}
	In this paper, we studied a special class of generalized Roth--Lempel
	codes
	\(\mathrm{GRL}_{k}(\boldsymbol{\alpha},\boldsymbol{v},
	\boldsymbol{A}_{3\times3})\)
	introduced by Li et al.~\cite{LFW}. We first derived a sufficient
	condition for
	\(\mathrm{GRL}_{k}(\boldsymbol{\alpha},\boldsymbol{v},
	\boldsymbol{A}_{3\times3})\) with $k\geq 5$
	to be non-GRS via Cauchy matrices. Using the necessary and
	sufficient condition for $\mathrm{GRL}_{k}(\boldsymbol{\alpha},\boldsymbol{v},
	\boldsymbol{A}_{3\times3})$ to be MDS and the parity-check matrix
	given by Li et al., we established necessary and sufficient conditions for $\mathrm{GRL}_{k}^{\perp}(\boldsymbol{\alpha},\boldsymbol{v},
	\boldsymbol{A}_{3\times3})$ to be AMDS and for $\mathrm{GRL}_{k}(\boldsymbol{\alpha},\boldsymbol{v},
	\boldsymbol{A}_{3\times3})$ to be self-dual.
	In particular, the criterion in Theorem \ref{evenselfdualGRLcode} does not depend on the value of \(q \pmod4\), extending the corresponding result in \cite{LFW}. We also used Schur square dimensions to distinguish these GRL codes from several known non-GRS code families under certain conditions. Future work may consider appended blocks \(\boldsymbol{A}_{{\color{red} l\times l}}\) with
	\({\color{red} l}\ge4\), and seek necessary and sufficient conditions for the corresponding codes to be
	non-GRS. Other directions include studying their weight distributions, hull dimensions, and related
	applications.

	\section{Declarations}
	\textbf{Acknowledgements} The authors thank the reviewers and the editor for their valuable comments and suggestions, which have improved the quality of this paper.
	
	\textbf{Availability of data and materials} Not applicable.
	
	\textbf{Competing interests} The authors declare no competing interests. 
	
	\textbf{Funding} This research was supported by the National Natural Science Foundation of China (Grant No. 12471494) and the Natural Science Foundation of Sichuan Province (2024NSFSC2051). 
	
	\textbf{Authors' contributions} Zhonghao Liang conceived the study and wrote the manuscript. Yongkang Wan and Qunying Liao checked the manuscript. All authors reviewed the manuscript.

\end{document}